\documentstyle[preprint,aps]{revtex}
\tightenlines

\newcommand{\nn}{\nonumber}
\def\dfrac#1#2{\displaystyle\frac{#1}{#2}}
\newcommand{\ovl}[1]{\overline{#1}}
\newcommand{\wt}[1]{\widetilde{#1}}

\newcommand{\eqn}[1]{(\ref{#1})}

\newcommand{\pslash}{p\kern-1ex /}
\newcommand{\lslash}{l\kern-1ex /}
\newcommand{\sslash}{s\kern-1ex /}
\newcommand{\Dslash}{{\cal D}\kern-1.5ex /}
\newcommand{\bpsi}{\overline{\psi}}
\newcommand{\bq}{{\overline{q}}}

\newcommand{\spr}{{s^\prime}}
\newcommand{\msbar}{{\overline {\rm MS}}}
\newcommand{\vev}[1]{\langle #1 \rangle}
\newcommand{\VEV}[3]{\left\langle #1\left| #2 \right| #3\right\rangle}
\newcommand{\beqa}{\begin{eqnarray}}
\newcommand{\eeqa}{\end{eqnarray}}

\begin{document}
\draft

\title{
\begin{flushright}
{\normalsize hep-lat/0206013}\\
{\normalsize UTHEP-457}\\
{\normalsize UTCCP-P-123}\\
\end{flushright}
Perturbative renormalization factors in domain-wall QCD with improved gauge 
actions}
\author{Sinya Aoki$^{1}$, 
Taku Izubuchi$^{2}$\thanks{on leave from Institute of Theoretical Physics, 
Kanazawa University,   Ishikawa 920-1192, Japan},
Yoshinobu Kuramashi$^{3}$ and Yusuke Taniguchi$^{1}$}
\address{
  ${}^{1}$Institute of Physics, University of Tsukuba,
  Tsukuba, Ibaraki 305-8571, Japan\\
  ${}^{2}$Physics Department, Brookhaven National Laboratory, Upton, NY 11973,
  USA\\
  ${}^{3}$High Energy Accelerator Research Organization (KEK),
  Tsukuba, Ibaraki 305-0801, Japan\\
}

\date{\today}

\maketitle

\begin{abstract}

We evaluate renormalization factors of the domain-wall fermion
system with various improved gauge actions at one loop level.
The renormalization factors are calculated for quark wave function,
quark mass, bilinear quark operators, three- and four-quark operators
in $\msbar$ scheme with the dimensional reduction(DRED) as well as 
the naive dimensional regularization(NDR).
We also present detailed results in the mean field improved perturbation 
theory.

\end{abstract}



\section{Introduction}

The domain wall fermion formalism
\cite{Kaplan92,Shamir93,Shamir95} offers a possibility of realizing
full chiral symmetry at finite lattice spacing that the explicit chiral
symmetry breaking term is suppressed exponentially in the fifth
dimensional length $N_5$.
This property is understood in terms of the overlap formalism\cite{NN94}
or the Ginsparg-Wilson relation\cite{Ginsparg,Luscher98}.

However realization of the exact chiral symmetry in $N_5\to\infty$ limit 
is non-trivial.
In Ref.~\cite{Shamir95} it is shown that the explicit breaking term in
the axial Ward-Takahashi identity vanishes exponentially in $N_5$ only
when the eigenvalues of the transfer matrix in the fifth direction are
strictly less than unity.
The exact chiral symmetry cannot be realized even in $N_5\to\infty$
limit if the largest eigenvalue of the transfer matrix becomes unity.
Recent studies
\cite{CV9812011,RBC9909117,CV9909140,cppacs99,cppacs-dwf,RBC0007038}
of the chiral properties in quenched domain-wall QCD (DWQCD)
seem to reveal that this is the case for the strong coupling region
around the lattice spacing $a^{-1}\sim1$ GeV.
By investigating the axial Ward-Takahashi identity it is found that
there remains non-zero chiral symmetry breaking term even in the
$N_5\to\infty$ limit\cite{cppacs-dwf,RBC0007038}
and such a residual quark mass extracted by the axial WT identity
becomes much larger than the physical $u,d$ quark masses at a reasonable
size of $N_5\sim20$ for numerical simulation.

The chiral property is improved in the weak coupling region around
$a^{-1}\sim2$ GeV.
The value of residual quark mass in the axial WT identity is much
smaller than in the strong coupling region.
However it is not still clear for the standard plaquette gauge action
whether chiral symmetry is broken slightly but
explicitly\cite{cppacs-dwf} or the symmetry breaking term vanishes 
exponentially in $N_5$ but decay rate is small\cite{RBC0007038}.
On the other hand, for the renormalization group (RG) improved gauge
action\cite{Iwasaki83} it was found in Ref.~\cite{cppacs-dwf} that
value of the residual mass is much smaller than that for the plaquette
action and furthermore the residual quark mass decays exponentially in
$N_5$ up to $N_5=24$, which
is consistent with a realization of the exact chiral symmetry.
We can conclude that chiral symmetry is much better realized with the RG
action than with the plaquette gauge action.
It is quite reasonable to adopt a combination of the domain-wall fermion 
and RG improved gauge action for computational simulation.

In this paper we evaluate the renormalization factors,
which is needed to convert the lattice quantities to the continuum ones, 
for the domain-wall fermion system with various improved gauge actions.
We calculate the renormalization factors of quark wave function, quark
mass, bilinear quark operators, three- and four-quark operators in
$\msbar$ scheme mainly with the dimensional reduction (DRED), 
and gives the relation between DRED and the naive dimensional regularization 
(NDR).
Since the domain-wall height $M$ (the mass in the five dimensional theory)
receives rather large additive quantum corrections, 
one must employ the mean field improved perturbation theory
for the reliable calculation of the renormalization factors.
We will explain this point in detail.

This paper is organized as follows.
In section~\ref{sec:action} we present the action and the corresponding
Feynman rules.
In section~\ref{sec:MF} we discuss the general form of the quantum
correction and introduce the mean field improvement in order to treat
the problem of the additive quantum correction to $M$.
Our main result is given in section~\ref{sec:renormalization}.
Finite part of the renormalization factor is evaluated numerically in
section~\ref{sec:numerical}.
We explain how to use the mean field improved results for 
various renormalization factors in section~\ref{sec:tool}.
We close the paper with a brief summary and comments 
in section~\ref{sec:conclusion}.

In this paper we take $SU(N)$ gauge
group with the gauge coupling $g$ and the second Casimir $C_F = 
\displaystyle \frac{N^2-1}{2N}$.
We set $N=3$ in the numerical calculations for three- and four-quark
renormalization factors.
The physical quantities are expressed in lattice units 
and the lattice spacing $a$ is suppressed unless necessary. 

\section{Action and Feynman rules}
\label{sec:action}

We employ the Shamir's domain-wall fermion action\cite{Shamir93,Shamir95}
given by
\begin{eqnarray}
S_f &=& \sum_{x,s,y,\spr} \bpsi(x,s) D_{dwf}(x,s;y,\spr) \psi(y,\spr)
 +\sum_x m \bq(x)  q(x)~,
\label{eq:action}
\\
D_{dwf}(x,s;y,\spr) &=& 
 D^4(x,y)\delta_{s,\spr} 
+D^5(s,\spr)\delta_{x,y}
+(M-5)\delta_{x,y} \delta_{s,\spr}~,
\label{eq:qm_dwf}
\\
D^4(x,y) &=& \sum_\mu
\frac{1}{2}\left[(1+\gamma_\mu)U_{x,\mu}\delta_{x+\hat\mu,y}
          + (1-\gamma_\mu)U^\dagger_{y,\mu}\delta_{x-\hat\mu,y}\right],
\\
D^5(s,\spr) &=& \left\{
\begin{array}{ll}
P_R\delta_{2,\spr}  & (s=1)\\
P_R\delta_{s+1,\spr} +P_L\delta_{s-1,\spr} & (1<s<N_5)\\
P_L\delta_{N_5-1,\spr} & (s=N_5) \\
\end{array}
\right. ,
\end{eqnarray}
where $x,y$ are four-dimensional space-time coordinates, and $s,s'$ are 
fifth-dimensional or ``flavor'' index,  bounded as 
$1 \le s, s' \le N_5$ with the free boundary condition at both ends.
In this paper we will take $N_5\to\infty$ limit and omit terms
suppressed exponentially in $N_5$.
$P_{R/L}$ is the projection matrix $P_{R/L}=(1\pm\gamma_5)/2$,
$m$ is physical quark mass and
the domain-wall height $M$ is a parameter of the theory which we set
$0\le M\le2$ in order to realize the massless fermion at tree level.
The quark mass term and quark operators for our calculation are
constructed with the 4-dimensional quark field defined on the edges of
the fifth dimensional space,
\begin{eqnarray}
q(x) = P_L \psi(x,1) + P_R \psi(x,{N_5}),
\nn \\
\ovl{q}(x) = \bpsi(x,{N_5}) P_L + \bpsi(x,1) P_R.
\label{eq:quark}
\end{eqnarray}

For the gauge part of the action we employ the following form in 4
dimensions:
\begin{equation}
S_{\rm gluon} = \frac{1}{g^2}\left\{
c_0 \sum_{\rm plaquette} {\rm Tr}U_{pl}
+ c_1  \sum_{\rm rectangle} {\rm Tr} U_{rtg}
+ c_2 \sum_{\rm chair} {\rm Tr} U_{chr}
+ c_3 \sum_{\rm parallelogram} {\rm Tr} U_{plg}\right\}, 
\label{eqn:RG}
\end{equation}
where the first term represents the standard plaquette action, and the 
remaining terms are six-link loops formed by a $1\times 2$ rectangle, 
a bent $1\times 2$ rectangle (chair) and a 3-dimensional parallelogram. 
The coefficients $c_0, \cdots, c_3$ satisfy the normalization condition
\begin{equation}
c_0+8c_1+16c_2+8c_3=1. 
\end{equation}
The RG-improved gauge action is defined by setting the parameters to the value 
suggested by an approximate renormalization group analysis.
In the following we will adopt the following choices 
$c_1=-0.331$, $c_2=c_3=0$(Iwasaki) and $c_1=-0.27, c_2+c_3=-0.04$(Iwasaki')
\cite{Iwasaki83}, $c_1=-0.252, c_2+c_3=-0.17$(Wilson)\cite{Wilson}
and $c_1 = -1.40686$, $c_2=c_3=0$(DBW2)\cite{qcdtaro} for the RG improved
gauge action, as well as $c_1=c_2=c_3=0$(Plaquette) and
$c_1=-1/12$, $c_2=c_3=0$(Symanzik)\cite{Weisz83,LW}.
With these choices of parameters the RG improved gauge action is expected to 
realize smooth gauge field fluctuations approximating those in the continuum 
limit better than with the unimproved plaquette action.

Weak coupling perturbation theory is developed by writing the link
variable in terms of the gauge potential
\begin{eqnarray}
U_{x,\mu}=\exp(igA_\mu(x+\frac{1}{2}\hat{\mu}))
\label{eqn:link}
\end{eqnarray}
and expanding in terms of the gauge coupling.
We adopt a covariant gauge fixing with a gauge parameter $\alpha$ 
defined by  
\begin{eqnarray}
S_{\rm GF} = \sum_x \frac{1}{2\alpha}
\left( \nabla_\mu A_\mu^a (x+\frac{1}{2}\hat{\mu}) \right)^2,
\end{eqnarray}
where $\nabla_\mu f_n\equiv (f_{n+\hat{\mu}}-f_n)$. 

The free part of the gluon action takes the form in momentum space
\begin{equation}
S_0 = \frac{1}{2} \int_{-\pi}^{\pi}\frac{d^4k}{(2\pi)^4}
\sum_{\mu, \nu}
A_\mu^a(k) 
\left(G_{\mu\nu}(k)
-\left(1-\frac{1}{\alpha}\right)\hat{k}_\mu\hat{k}_\nu\right)
A_\nu^a(-k) , 
\end{equation}
where
\begin{equation}
G_{\mu \nu}(k) = \hat{k}_\mu \hat{k}_\nu + \sum_\rho 
(\hat{k}_\rho \delta_{\mu \nu} - \hat{k}_\mu \delta_{\rho
\nu}) q_{\mu \rho} \hat{k}_\rho
\end{equation}
with 
\begin{equation}
\hat{k}_\mu = 2 \sin\frac{k_\mu}{2}
\end{equation}
and $q_{\mu \nu}$ is defined as
\begin{equation}
q_{\mu \nu} = (1-\delta_{\mu\nu})\left(1 - (c_1 - c_2 - c_3)
(\hat{k}_\mu^2 + \hat{k}_\nu^2) -(c_2 + c_3) \hat{k}^2\right) .
\end{equation}
The gluon propagator can be written as 
\begin{eqnarray}
D_{\mu \nu}(k) &=& (\hat{k}^2)^{-2} \left[
\hat{k}_\mu \hat{k}_\nu  + \sum_\sigma 
(\hat{k}_\sigma \delta_{\mu \nu} - \hat{k}_\nu \delta_{\mu
\sigma} ) \hat{k}_\sigma A_{\sigma \nu} \right]
-(1-\alpha) \frac{\hat{k}_\mu \hat{k}_\nu}{(\hat{k}^2)^2}
\\
& = & (\hat{k}^2)^{-2} \left[
(1 - A_{\mu \nu} )\hat{k}_\mu \hat{k}_\nu 
+ \delta_{\mu \nu} \sum_\sigma \hat{k}_\sigma^2
A_{\nu \sigma} \right]
-(1-\alpha) \frac{\hat{k}_\mu \hat{k}_\nu}{(\hat{k}^2)^2} , 
\end{eqnarray}
where $A_{\mu \nu}$ is a function of $q_{\mu \nu}$ and
$\hat{k}_\mu$ whose form we refer to the original 
literatures\cite{Iwasaki83,Weisz83}. 
In this paper we will adopt the Feynman gauge($\alpha=1$) without loss of
generality, since the renormalization factors for the physical quantities 
such as quark mass, bilinear quark operators, three- and four-quark 
operators do not depend on the choice of the gauge fixing condition.

For the fermion part we need following three types of propagators in
domain-wall QCD.
One is the propagator which connects general flavor indices,
\begin{eqnarray}
S(p)_{st} &=&
\sum_{u=1}^{\infty}
\left( -i\gamma_\mu \sin p_\mu + W^- + m M^- \right)_{su} G_R(u,t) P_R
\nn\\&+&
\sum_{u=1}^{\infty}
\left( -i\gamma_\mu \sin p_\mu + W^+ + m M^+ \right)_{su} G_L(u,t) P_L,
\end{eqnarray}
where the mass matrix is
\begin{eqnarray}
&&
W^{+} =
\pmatrix{
-W & 1  &        &    \cr
   & -W & \ddots &    \cr
   &    & \ddots & 1  \cr
   &    &        & -W \cr
},
\quad
W^{-}_{s,t} =
\pmatrix{
-W &        &        &    \cr
1  & -W     &        &    \cr
   & \ddots & \ddots &    \cr
   &        & 1      & -W \cr
},
\label{eqn:mass-matrix-p2}
\\&&
M^+ = \pmatrix{
  &  &  \cr
  &  &  \cr
1 &  &  \cr}
,\quad
M^- = \pmatrix{
  &  & 1\cr
  &  &  \cr
  &  &  \cr}
\label{eqn:mass-matrix-m2}
\end{eqnarray}
and $G_{R/L}$ is given by 
\begin{eqnarray}
G_{R} (s, t) 
&=&
\frac{A}{F}
\Bigl[
-(1-m^2) \left(1-W e^{-\alpha}\right) e^{\alpha (-2N+s+t)}
-(1-m^2) \left(1-W e^\alpha\right) e^{-\alpha (s+t)}
\nn\\&&
-2W \sinh (\alpha) m
 \left( e^{\alpha (-N+s-t)} + e^{\alpha (-N-s+t)} \right)
\Bigr]
+ A e^{-\alpha |s-t|},
\\
G_{L} (s, t) 
&=&
\frac{A}{F}
\Bigl[
-(1-m^2) \left(1-We^{\alpha}\right) e^{\alpha (-2N+s+t-2)}
-(1-m^2) \left(1-We^{-\alpha}\right) e^{\alpha (-s-t+2)}
\nn\\&&
-2W \sinh (\alpha) m
 \left( e^{\alpha (-N+s-t)} +  e^{\alpha (-N-s+t)} \right)
\Bigr]
+ A e^{-\alpha |s-t|},
\\
\cosh (\alpha) &=& \frac{1+W^2+\sum_\mu \sin^2 p_\mu}{2|W|},
\\
A&=& \frac{1}{2W \sinh (\alpha)},
\\
F &=& 1-e^{\alpha} W-m^2 \left(1-W e^{-\alpha}\right),
\\
W &=& 1-M +\sum_\mu (1-\cos p_\mu).
\end{eqnarray}
When $W$ becomes negative the fermion propagator is given with the
replacement $e^{\pm\alpha}\to-e^{\pm\alpha}$.

The second one is the propagator which connects the physical quark
field and the fermion field of general flavor index,
\begin{eqnarray}
&&
\vev{q(p) \bpsi(-p,s)} 
=
\frac{1}{F}
\left( i\gamma_\mu \sin p_\mu -m\left(1 -W e^{-\alpha} \right)
\right)
\left( e^{-\alpha (N-s)} P_R + e^{-\alpha (s-1)} P_L \right)
\nn\\&&\qquad
+\frac{1}{F} \Bigl[
m\left(i\gamma_\mu \sin p_\mu -m\left(1-W e^{-\alpha}\right)\right)
- F \Bigr] e^{-\alpha}
\left( e^{-\alpha (s-1)} P_R + e^{-\alpha (N-s)} P_L \right),
\\&&
\vev{\psi(p,s) \ovl{q}(-p)} 
=
\frac{1}{F}
\left( e^{-\alpha (N-s)} P_L + e^{-\alpha (s-1)} P_R \right)
\left( i\gamma_\mu \sin p_\mu -m\left(1 - W e^{-\alpha} \right)
\right)
\nn\\&&\qquad
+\frac{1}{F}
\left( e^{-\alpha (s-1)} P_L + e^{-\alpha (N-s)} P_R \right) e^{-\alpha}
\Bigl[
m\left(i\gamma_\mu \sin p_\mu -m\left(1- We^{-\alpha}\right) \right)
- F \Bigr]
\end{eqnarray}
and the third one is two point function of the physical quark field
\begin{eqnarray}
S_q(p) \equiv \vev{q(p) \ovl{q}(-p)} = 
 \frac{-i\gamma_\mu \sin p_\mu + \left(1-W e^{-\alpha}\right) m}
{-\left(1-e^{\alpha}W\right) + m^2 (1-W e^{-\alpha})}.
\end{eqnarray}
In the continuum limit the physical quark propagator becomes
\begin{eqnarray}
S_q(p) &=&
\frac{(1-w_0^2)}{i\pslash + (1-w_0^2) m}.
\end{eqnarray}
Here we notice that the overall factor $1-w_0^2$ with $w_0=1-M$ appears
in the quark wave function and mass.

Following two interaction vertices will concern in the one loop correction,
\begin{eqnarray}
V_{1\mu}^a(k,p)_{st}
&=& -i g T^a \left(\gamma_\mu\ovl{V}_{1\mu}(k,p)+\wt{V}_{1\mu}(k,p)\right)
\delta_{st} ,
\\
V_{2\mu\nu}^{ab}(k,p)_{st}
&=& \frac{1}{2} g^2 \frac{1}{2}\{T^{a}, T^{b}\}
\left(\gamma_\mu\wt{V}_{1\mu}(k,p)+\ovl{V}_{1\mu}(k,p)\right)\delta_{\mu\nu}
\delta_{st} ,
\end{eqnarray}
where
\begin{eqnarray}
\ovl{V}_{1\mu}(k,p) &=& \cos \frac{1}{2}(-k_\mu + p_\mu),
\\
 \wt{V}_{1\mu}(k,p) &=& i\sin \frac{1}{2}(-k_\mu + p_\mu).
\end{eqnarray}

In this paper we will calculate the one loop correction to the Green
functions constructed with physical quark fields $q, \bq$.
When we carry out the perturbative calculation we first notice that the
external line propagator is expanded in terms of the external momentum
$p_\mu$ and quark mass $m$ as
\begin{eqnarray}
&&
\vev{q(p) \bpsi(-p,s)}
\to
\frac{1-w_0^2}{i\pslash +(1-w_0^2)m}
\left(L(s)-\frac{w_0 i\pslash}{1-w_0^2}R(s)\right),
\\&&
\vev{\psi(p,s) \ovl{q}(-p)} 
\to
\left(R(s)-L(s)\frac{w_0 i\pslash}{1-w_0^2}\right)
\frac{1-w_0^2}{i\pslash + (1-w_0^2) m},
\end{eqnarray}
where
\begin{eqnarray}
L(s) &=& \left( w_0^{(N-s)} P_R + w_0^{(s-1)} P_L \right),
\\
R(s) &=& \left( w_0^{(s-1)} P_R + w_0^{(N-s)} P_L \right).
\end{eqnarray}
The fifth dimensional index $s,t$ are summed with this $L(s), R(s)$ and
following form of the propagators will concern in the loop integral,
\begin{eqnarray}
S_{RR}(p) &\equiv& \sum_{s,t=1}^\infty R(s)S(p)_{st}R(t)
\nn\\&=&
-i \gamma_\mu \sin p_\mu 
\frac{1}{F}\left(-\frac{m e^{-\alpha}}{(1-w_0 e^{-\alpha})^2}\right)
 + (w_0-W)\wt{G_R}
 -\frac{1}{F}\frac{m^2 e^{-\alpha}}{1-w_0e^{-\alpha}},
\label{eqn:sum1}
\\
S_{LL}(p) &\equiv& \sum_{s,t=1}^\infty L(s)S(p)_{st}L(t) = S_{RR}(p),
\\
S_{RL}(p) &\equiv& \sum_{s,t=1}^\infty R(s)S(p)_{st}L(t)
=
-i \gamma_\mu \sin p_\mu \wt{G}_L
-\frac{m}{F}\frac{1-We^{-\alpha}}{(1-w_0 e^{-\alpha})^2},
\\
S_{LR}(p) &\equiv& \sum_{s,t=1}^\infty L(s)S(p)_{st}R(t)
=
-i\gamma_\mu \sin p_\mu \wt{G}_R
-\frac{m}{F}\frac{e^{-2\alpha}\left(1-We^{\alpha}\right)}
 {(1-w_0 e^{-\alpha})^2},
\\
S_{\rm Lq}(p) &\equiv& \sum_{s=1}^\infty L(s) \vev{\psi(p,s)\ovl{q}(-p)}
=
\left(\frac{e^{-\alpha}}{F(1-w_0 e^{-\alpha})} \right)
\biggl(im\gamma_\mu \sin p_\mu -\left(1-We^{\alpha}\right)\biggr),
\\
S_{\rm qR}(p) &\equiv& \sum_{s=1}^\infty \vev{q(p)\bpsi(-p,s)} R(s)
=S_{\rm Lq}(p),
\\
S_{\rm qL}(p) &\equiv& \sum_{s=1}^\infty \vev{q(p)\bpsi(-p,s)} L(s)
=
\frac{1}{1-w_0 e^{-\alpha}} \frac{1}{F}
\left(i\gamma_\mu \sin p_\mu - m\left(1-W(p) e^{-\alpha}\right) \right),
\\
S_{\rm Rq}(p) &\equiv& \sum_{s=1}^\infty R(s) \vev{\psi(p,s)\ovl{q}(-p)}
=S_{\rm qL}(p),
\label{eqn:sum8}
\end{eqnarray}
where
\begin{eqnarray}
\wt{G_L} &=&
\frac{1}{2W\sinh\alpha}
\Biggl[
\frac{\sinh\alpha_0 -\sinh\alpha}
 {2w_0\sinh\alpha_0 (\cosh\alpha_0-\cosh\alpha)}
-(1-m^2)\frac{1}{F}\frac{1-We^{-\alpha}}{(1-w_0e^{-\alpha})^2}
\Biggr],
\\
\wt{G_R} &=&
\frac{1}{2W\sinh\alpha}
\Biggl[
\frac{\sinh\alpha_0 -\sinh\alpha}
 {2w_0\sinh\alpha_0 (\cosh\alpha_0-\cosh\alpha)}
-(1-m^2)\frac{1}{F}\frac{e^{-2\alpha}(1-We^{\alpha})}{(1-w_0e^{-\alpha})^2}
\Biggr].
\end{eqnarray}

\section{Mean field improvement}
\label{sec:MF}

Before evaluating the one loop correction we will discuss on the mean
field improvement in this section.
In the domain wall formalism the renormalization factor of an $n$-quark 
operator $O_n$ has a generic form
\begin{eqnarray}
&&
O_n^{\ovl{\rm MS}}(\mu) = Z O_n^{\rm lattice}(1/a),
\\&&
Z=(1-w_0^2)^{-n/2}Z_w^{-n/2}Z_{O_n},
\end{eqnarray}
where $Z_w$ represents the quantum correction to the 
normalization factor $1-w_0^2$ of physical quark fields $q, \bq$, 
and $Z_{O_n}$ is the vertex correction to $O_n$.
Here we notice that $Z_w$ is written in the form
\begin{eqnarray}
Z_w = 1+\frac{2w_0}{1-w_0^2} \frac{g^2 C_F}{16\pi^2} \Sigma_w.
\end{eqnarray}
As is known in perturbative calculation \cite{zfactor}
the one-loop correction in $Z_w$ becomes huge for some choice of
$M$ because of the tadpole contribution in $\Sigma_w$ and division with
$1-w_0^2$.
This reflects the fact that the one loop correction to domain-wall
height $M$ is additive rather than multiplicative\cite{Aoki-Taniguchi}.
Rewriting it to the multiplicative form
\begin{eqnarray}
1-\left(w_0-\frac{g^2 C_F}{16\pi^2} \Sigma_w\right)^2
\to \left(1-w_0^2\right) Z_w
\end{eqnarray}
can be done only when $g^2\ll1$ since the correction $\Sigma_w$ contains
the tadpole contribution and becomes large\cite{zfactor}.

To carry out this rewriting reliably, we adopt the mean field improvement
as follows.
\begin{eqnarray}
1-\left(w_0-\frac{g^2 C_F}{16\pi^2} \Sigma_w\right)^2
&=&
1-\left(w_0+g^2C_F 2T_{\rm MF}
 +\frac{g^2 C_F}{16\pi^2}\left(-\Sigma_w-16\pi^2 2T_{\rm MF}\right)\right)^2
\nn\\&\to&
1-\left(w_0+4(1-u)
 +\frac{g^2 C_F}{16\pi^2}\left(-\Sigma_w-16\pi^2 2T_{\rm MF}\right)\right)^2
\nn\\&=&
1-\left(w_0^{\rm MF}
 +\frac{g^2 C_F}{16\pi^2}\left(-\Sigma_w-16\pi^2 2T_{\rm MF}\right)\right)^2
\nn\\&=&
\left(1-\left(w_0^{\rm MF}\right)^2\right)
\left(
1+\frac{2w_0^{\rm MF}}{1-\left(w_0^{\rm MF}\right)^2}\frac{g^2 C_F}{16\pi^2}
 \left(\Sigma_w+16\pi^2 2T_{\rm MF}\right)
\right)
\nn\\&=&
\left(1-\left(w_0^{\rm MF}\right)^2\right) Z_w^{\rm MF},
\label{eqn:additive}
\end{eqnarray}
where $T_{\rm MF}$ is the one-loop correction to the mean-field factor 
defined by
\begin{eqnarray}
u=1-g^2C_F\frac{T_{\rm MF}}{2}+\cdots.
\end{eqnarray}
where $u=P^{1/4}$ with $P$ being the plaquette.
In the second line of equation \eqn{eqn:additive} we have replaced the
perturbative correction $g^2C_FT_{\rm MF}$ to the domain-wall height with
nonperturbative value, $2(1-u)$, according to the standard
procedure of the mean field improved perturbation theory.
On the other hand,
the perturbative value is still used for $Z_w^{\rm MF}$,
since the mean field improved value, $\Sigma_w+16\pi^2 2T_{\rm MF}$, 
becomes small enough.

The procedure of the mean field improvement for the renormalization factor of 
the general $n$-quark operator $O_n$ becomes as follows.
We factor out the mean field contribution perturbatively from
the vertex correction $Z_{O_n}$ and replace it with the non-perturbative one:
\begin{eqnarray}
Z_{O_n} &=&
\left(1-g^2C_F\frac{n}{4}T_{\rm MF}\right)
\left(Z_{O_n}+g^2C_F\frac{n}{4}T_{\rm MF}\right)
\to
u^{n/2}\left(Z_{O_n}+g^2C_F\frac{n}{4}T_{\rm MF}\right),
\end{eqnarray}
where $u=P^{1/4}$ is evaluated numerically.
This leads to the rewriting of the total renormalization factor of $O_n$, 
\begin{eqnarray}
Z \to Z^{\rm MF} =
\left(1-\left(w_0^{\rm MF}\right)^2\right)^{-n/2}
\left(Z_w^{\rm MF}\right)^{-n/2}
 u^{n/2} Z_{O_n}^{\rm MF},
\end{eqnarray}
where
\begin{eqnarray}
&&
w_0^{\rm MF} = w_0+4(1-u),
\label{eqn:mfM}
\\&&
Z_w^{\rm MF}= Z_w|_{w_0=w_0^{\rm MF}}
+\frac{4w_0^{\rm MF}}{1-(w_0^{\rm MF})^2}g^2C_FT_{\rm MF}.
\end{eqnarray}
Note that the difference between the mean field improved renormalization factor
and the unimproved one is of higher order in the perturbative expansion.
The renormalization factors of the quark wave function, the quark mass and
the $n$-quark operator are shifted as
\begin{eqnarray}
&&
Z_2^{\rm MF} =
Z_2|_{w_0=w_0^{\rm MF}}+\frac{1}{2}g^2C_FT_{\rm MF},
\\&&
Z_m^{\rm MF} =
Z_m|_{w_0=w_0^{\rm MF}}-\frac{1}{2}g^2C_FT_{\rm MF},
\\&&
Z_{O_n}^{\rm MF} =
Z_{O_n}|_{w_0=w_0^{\rm MF}}+\frac{n}{4}g^2C_FT_{\rm MF}.
\end{eqnarray}

Now we have a short comment on derivation of $T_{\rm MF}$.
The one loop correction is given by expanding the plaquette value in
gauge coupling and executing the momentum integral.
\begin{eqnarray}
P=1-g^2C_F2T_{\rm MF}=1-g^2C_F2\left(T+\delta T\right),
\label{eqn:TMF}
\end{eqnarray}
where $T$ is a tadpole contribution and $\delta T$ is a remaining
contribution,
\begin{eqnarray}
T &=& \int_k D_{\mu\mu}(k),
\label{eqn:tadpole}
\\
\delta T &=& -\int_k
\left(
\frac{1}{2}\left(
\cos k_\nu D_{\mu\mu}(k)+\cos k_\mu D_{\nu\nu}(k)
\right)
+2\sin\frac{k_\mu}{2}\sin\frac{k_\nu}{2}D_{\mu\nu}(k)
\right)
\end{eqnarray}
with $\mu$, $\nu$ are unsummed and $\mu\neq\nu$. 
Here momentum integral means
\begin{eqnarray}
\int_k=\int_{-\pi}^\pi\frac{d^4k}{(2\pi)^4}.
\end{eqnarray}
After numerical integration we get
\begin{eqnarray}
T_{\rm MF}=\cases{
1/8 & ({\rm Plaquette})\cr
0.0525664 & ({\rm Iwasaki, $c_1=-0.331, c_1+c_2=0$})\cr
0.0191580 & ({\rm DBW2, $c_1=-1.40686, c_1+c_2=0$})\cr
0.0915657 & ({\rm Symanzik, $c_1=-1/12, c_2+c_3=0$})\cr
0.0552016 & ({\rm Iwasaki', $c_1=-0.27, c_2+c_3=-0.04$})\cr
0.0482425 & ({\rm Wilson, $c_1=-0.252, c_2+c_3=-0.17$})}.
\end{eqnarray}
Here please notice that the finite part $\Sigma_w$ and the tadpole
factor $T_{\rm MF}$ which is defined through
plaquette $P$ are gauge independent.

The mean field improved factor can also defined with perturbative
evaluation of the link variable \eqn{eqn:link}.
In this case the averaged link variable is written in terms of $T$ and gauge
dependent part $\delta T_{\rm gauge}$,
\begin{eqnarray}
u=\vev{U_\mu(n)}
=1-C_Fg^2 \frac{1}{2}\left(T-(1-\alpha)\delta T_{\rm gauge}\right).
\end{eqnarray}
The gauge dependent term  is independent on the choice of gauge
action,
\begin{eqnarray}
\delta T_{\rm gauge} = 0.0387334.
\end{eqnarray}
$T$ is given by
\begin{eqnarray}
T=\cases{
0.1549334 & ({\rm Plaquette})\cr
0.0947597 & ({\rm Iwasaki, $c_1=-0.331, c_1+c_2=0$})\cr
0.0624262 & ({\rm DBW2, $c_1=-1.40686, c_1+c_2=0$})\cr
0.1282908 & ({\rm Symanzik, $c_1=-1/12, c_2+c_3=0$})\cr
0.0973746 & ({\rm Iwasaki', $c_1=-0.27, c_2+c_3=-0.04$})\cr
0.0916234 & ({\rm Wilson, $c_1=-0.252, c_2+c_3=-0.17$})}.
\end{eqnarray}
As the third choice, one may define the mean field improved factor $u$
through the critical hopping parameter $K_c$ of the Wilson fermion
action as
\begin{eqnarray}
u &=&\frac{1}{8 K_c}
=1-C_Fg^2 \frac{1}{2} T_{K_c},
\end{eqnarray}
where\cite{ANTU}
\begin{eqnarray}
T_{K_c}=\cases{
0.162858 & ({\rm Plaquette})\cr
0.082555 & ({\rm Iwasaki, $c_1=-0.331, c_1+c_2=0$})\cr
-0.036482& ({\rm DBW2, $c_1=-1.40686, c_1+c_2=0$})\cr
0.128057 & ({\rm Symanzik, $c_1=-1/12, c_2+c_3=0$})\cr
0.086167 & ({\rm Iwasaki', $c_1=-0.27, c_2+c_3=-0.04$})\cr
0.078169 & ({\rm Wilson, $c_1=-0.252, c_2+c_3=-0.17$})}.
\end{eqnarray}

\section{One loop renormalization factors}
\label{sec:renormalization}

In this section we derive the renormalization factors at the one loop level 
for the quark wave function, the quark mass, bilinear quark operators, 
three- and four- quark operators in the form of momentum integrals.
The loop integral will be carried out numerically in the next section.
Matching of the lattice and continuum operators are carried out at a
scale $\mu$ in the $\msbar$ scheme with the dimensional reduction (DRED)
or the naive dimensional regularization (NDR).
Difference between NDR and DRED resides only in the finite parts of the
renormalization factors in the continuum.
The finite parts on lattice are derived with mean field improvement.
Hereafter we suppress the index MF in quantities unless confusion may arise.

\subsection{Quark propagator}

One loop correction to the physical quark propagator is given by two
diagrams.
Contribution from the tadpole diagram is given by
\begin{eqnarray}
G_{\rm tad} &=& 
\sum_{s=1}^{\infty}
\frac{1-w_0^2}{i\pslash +(1-w_0^2)m}
\left(L(s)-\frac{w_0 i\pslash}{1-w_0^2}R(s)\right)
V_{2\mu\nu}^{ab}(-p,p)
\delta^{ab}\int_{k}D_{\mu\nu}(k)
\nn\\&&\times
\left(R(s)-L(s)\frac{w_0 i\pslash}{1-w_0^2}\right)
\frac{1-w_0^2}{i\pslash + (1-w_0^2) m}.
\end{eqnarray}
Contribution from the rising sun diagram is written as
\begin{eqnarray}
G_{\rm rs} &=&
\sum_{s,t=1}^{\infty}
\frac{1-w_0^2}{i\pslash +(1-w_0^2)m}
\left(L(s)-\frac{w_0 i\pslash}{1-w_0^2}R(s)\right)
\nn\\&&\times
\int_{k} V_{1\mu}^a(-p,p-k) S(p-k)_{st} V_{1\nu}^b(-(p-k),p)
\delta^{ab}D_{\mu\nu}(k)
\nn\\&&\times
\left(R(t)-L(t)\frac{w_0 i\pslash}{1-w_0^2}\right)
\frac{1-w_0^2}{i\pslash + (1-w_0^2) m}.
\end{eqnarray}
Making use of the summation formula
\eqn{eqn:sum1}-\eqn{eqn:sum8} and following the calculation
in Ref.~\cite{zfactor}, we obtain the ``full'' quark propagator
at the one loop level  on lattice and in the continuum with $\msbar$ scheme 
as follows.
\begin{eqnarray}
\langle q_{\rm lat} \bar q_{\rm lat}\rangle
&=&\dfrac{(1-w_0^2)Z_w u^{-1} Z_2^{\rm lat}}
{i\pslash +(1-w_0^2)Z_w m_{\rm lat}/(u Z_m^{\rm lat})},
\\
\langle q_{\ovl{\rm MS}} \bar q_{\ovl{\rm MS}}\rangle
&=&\dfrac{Z_2^{\ovl{\rm MS}}}
{i\pslash+m_{\ovl{\rm MS}}/Z_m^{\ovl{\rm MS}}},
\end{eqnarray}
where the mean field improvement is used on lattice.
Comparing these two expression for the quark propagator, we obtain
the following relations.
\begin{eqnarray}
q_{\ovl{\rm MS}} &=& (1-w_0^2)^{-1/2} Z_w^{-1/2} (u Z_2)^{1/2} q_{\rm lat},
\\
m_{\ovl{\rm MS}} &=& (1-w_0^2) Z_w Z_m u^{-1} m_{\rm lat},
\end{eqnarray}
where
\begin{eqnarray}
Z_2 &=& Z_2^{\ovl{\rm MS}}/Z_2^{\rm lat},
\\
Z_m &=& Z_m^{\ovl{\rm MS}}/Z_m^{\rm lat}.
\end{eqnarray}
The explicit form for $Z$  factors are given below.
The renormalization factor for $1-w_0^2$ is written as
\begin{eqnarray}
Z_w &=& 1 + \dfrac{g^2}{16\pi^2} C_F z_w^{\rm MF}
\\
z_w^{\rm MF} &=& \dfrac{2w_0}{1-w_0^2}
 \left[\Sigma_w+32\pi^2T_{\rm MF}\right], 
\\
\Sigma_w &=& 16\pi^2\left\{ -2T + 4(1-w_0^2)\int_k\dfrac{(T_4 -T_3)
S_F +T_1 (\wt{G_L}|_{m=0} +\wt{G_R}|_{m=0})}{G_0^2} \right\}.
\end{eqnarray}
The quark wave function renormalization factor is given as
\begin{eqnarray}
Z_2 &=& 1 +\dfrac{g^2}{16\pi^2}C_F
\left[- \log (\mu a)^2 + z_2^{\rm MF}\right],
\\
z_2^{\rm MF} &=& \Sigma_1^{\ovl{\rm MS}}-\Sigma_1
 + 16\pi^2\dfrac{ T_{\rm MF}}{2},
\\
\Sigma_1^{\ovl{\rm MS}} &=& -\frac{1}{2} ({\rm DRED}),
 \frac{1}{2} ({\rm NDR}),
\\
\Sigma_1 &=& 16\pi^2\dfrac{T}{2} + 16\pi^2
\int_k \Biggl\{ \dfrac{1-w_0^2}{G_0^2}
\Bigl[ \wt{G_L}|_{m=0} ( -2 T_1 + 2 T_5 - c T_4)
\nn\\&+&
 \wt{G_R}|_{m=0}(-2(w_0-W)(T_4-T_3)- c T_3)
\nn\\&+&
 f_L^\beta (2 T_2^\beta - 4 T_4 s_\beta^2 )
+ f_R^\beta (-2(w_0-W) T_1^\beta - 4 s_\beta^2)
\Bigr] + \dfrac{\theta (\pi^2-k^2)}{(k^2)^2}\Biggr\} 
-\log \pi^2.
\end{eqnarray}
The quark mass renormalization factor becomes
\begin{eqnarray}
Z_m &=& 1+\dfrac{g^2}{16\pi^2}C_F\left[-3\log(\mu a)^2 + z_m^{\rm MF}\right],
\\
z_m^{\rm MF} &=& (\Sigma_2^{\ovl{\rm MS}}-\Sigma_2)
 -(\Sigma_1^{\ovl{\rm MS}}-\Sigma_1)-16\pi^2\dfrac{T_{\rm MF}}{2},
\\
\Sigma_2^{\ovl{\rm MS}}&=& -4({\rm DRED}), -2({\rm NDR}),
\\
\Sigma_2 &=& 4\times 16\pi^2 \int_k \left\{ \dfrac{1}{G_0^2}
\left[ T_4 F_m^{RL} + 2 T_1 F_m - T_3 F_m^{LR}\right]
 + \dfrac{\theta (\pi^2-k^2)}{(k^2)^2}\right\} 
-4\log\pi^2.
\end{eqnarray}
Here $T$ is a tadpole factor \eqn{eqn:tadpole} and $T_{\rm MF}$ is given 
in \eqn{eqn:TMF}.
No sum is taken for $\beta$.
We have used the following short-handed notations;
$s_\mu =\sin k_\mu$, $c_\mu =\cos k_\mu$,
$\hat s_\mu = \sin \dfrac{k_\mu}{2}$, 
$\hat c_\mu = \cos \dfrac{k_\mu}{2}$, 
$ c = \sum_\mu c_\mu $,
$G_0=\hat{k}^2$, $s^2 =\sum_\mu s_\mu^2$, $\hat s^2 =\sum_\mu \hat s_\mu^2$, 
\begin{eqnarray}
\wt{ G_L}|_{m=0}
 &=&\dfrac{1}{2 W sh }\left[\dfrac{sh_0-sh}{2 w_0 sh_0(ch_0-ch)}
-\dfrac{1-W e^{-\alpha}}{F_0 (1-w_0e^{-\alpha})^2}\right],
\\
\wt{ G_R}|_{m=0}
 &=&\dfrac{1}{2 W sh }\left[\dfrac{sh_0-sh}{2 w_0 sh_0(ch_0-ch)}
-\dfrac{e^{-2\alpha}}{(1-w_0e^{-\alpha})^2}\right],
\\
S_F &=& (w_0 - W) \wt{ G_R}|_{m=0},
\\
f_L^\beta &=& \left(\dfrac{r}{W}-\dfrac{ch}{sh}g_\beta \right)
 \wt{ G_L}|_{m=0}
+\dfrac{g_\beta}{2W sh}\dfrac{1}{2 w_0 sh_0}
\dfrac{1+sh_0 sh - ch_0 ch}{(ch_0-ch)^2}
+\dfrac{1}{W sh}\dfrac{1}{F^2|_{m=0}}
\nn \\
&\times&\dfrac{r\cdot sh - W g_\beta (ch-W)}{(1-w_0e^{-\alpha})^2}
+\dfrac{g_\beta}{W sh}\dfrac{1-We^{-\alpha}}{F_0}
\dfrac{w_0 e^{-\alpha}}{(1-w_0e^{-\alpha})^3},
\\
g_\beta &=& \dfrac{c_\beta+r (ch-W)}{ W\ sh },
\\
f_R^\beta &=& \left(\dfrac{r}{W}-\dfrac{ch}{sh}g_\beta \right)
 \wt{ G_R}|_{m=0}
+\dfrac{g_\beta}{2W sh}\dfrac{1}{2 w_0 sh_0}
\dfrac{1+sh_0 sh - ch_0 ch}{(ch_0-ch)^2}
+\dfrac{1}{W sh}\dfrac{g_\beta e^{-2\alpha}}{(1-w_0e^{-\alpha})^3},
\nn\\
\\
F^\beta &=& ( w_0 - W) f_R^\beta + r \wt{ G_R}|_{m=0},
\\
F_m &=&\dfrac{e^{-\alpha}}{F_0 (1-w_0e^{-\alpha})^2},
\\
F_m^{RL}&=&\dfrac{1-We^{-\alpha}}{F_0}\dfrac{1}{(1-w_0e^{-\alpha})^2},
\\
F_0&=&1-We^\alpha,
\\
F_m^{LR}&=& \dfrac{e^{-2\alpha}}{(1-w_0e^{-\alpha})^2},
\\
T_1 &=& \dfrac{1}{2} \hat s^2  s^2,
\\
T_1^\beta &=& = 2 \hat s^2 s_\beta^2,
\\
T_2^\beta &=& s^2 s_\beta^2 
+ 4 s_\beta^2 \hat c_\beta^2 \sum_\mu\bar A_{\beta\mu}\hat s_\mu^2 
- s_\beta^2 \sum_\mu\bar A_{\beta\mu} s_\mu^2,
\\
T_3 &=& (\hat s^2)^2,
\\
T_4 &=&\sum_{\mu\nu} \hat c_\mu^2 \bar A_{\mu\nu} \hat s_\nu^2,
\\
T_5 &=& \sum_{\mu\nu} c_\mu \hat c_\mu^2 \bar A_{\mu\nu} \hat s_\nu^2,
\\
\bar A_{\mu\nu} &=& \delta_{\mu\nu}+A_{\mu\nu}.
\end{eqnarray}
Some fundamental quantities are given by
\begin{eqnarray}
W &=& 1-M+4(1-u) - r\sum_\mu ( 1- c_\mu), \quad w_0 = 1-M+4(1-u),
\\
ch &=& \cosh(\alpha) = \dfrac{ 1+ W^2+s^2}{2 W },\quad
ch_0 = \cosh(\alpha_0)=\dfrac{ 1+ w_0}{2 w_0 },
\\
sh &=& \sinh(\alpha), \quad sh_0 = \sinh(\alpha_0) .
\end{eqnarray}
Here we notice that the domain-wall height is shifted with
nonperturbative mean field improvement factor $u$ in this section, 
which is essential for $z_w^{\rm MF}$ as is seen in the previous section.

\subsection{Bilinear operators}

We consider the local bilinear operators constructed with physical quark 
fields,
\begin{eqnarray}
O_\Gamma = \bq \Gamma q,
\end{eqnarray}
where
\begin{eqnarray}
\Gamma=1(S),\gamma_5(P),\gamma_\mu(V),\gamma_\mu\gamma_5(A)
 ,\sigma_{\mu\nu}(T).
\end{eqnarray}
One loop correction to the bilinear quark operators is evaluated in the
same way as in Ref.~\cite{zfactor}.
The operator matching relation is given by
\begin{eqnarray}
O_\Gamma^{\ovl{MS}}(\mu) &=& (1-w_0^2)^{-1}Z_w^{-1} u Z_\Gamma(\mu a) 
O_{\Gamma}^{\rm lat}(1/a),
\end{eqnarray}
where the renormalization factor is given by
\begin{eqnarray}
&&
Z_\Gamma = 1 + \dfrac{g^2}{16\pi^2} C_F\left[
\left(\dfrac{h_2(\Gamma)}{4}-1\right)\log (\mu a)^2
 + z_\Gamma^{\rm MF}\right],
\\&&
z_\Gamma^{\rm MF} = z_\Gamma^{\ovl{\rm MS}}
 -z_\Gamma^{\rm lat} +16\pi^2\dfrac{T_{\rm MF}}{2},
\label{eqn:zb}
\\&&
h_2(\Gamma)=4({\rm VA}), 16({\rm SP}), 0({\rm T}),
\\&&
z_\Gamma^{\ovl{\rm MS}}\mbox{ (DRED) }
= V_\Gamma^{\ovl{\rm MS}}-1/2 = 1/2\mbox{ (VA) }, 
7/2\mbox{ (SP) }, -1/2\mbox{ (T) },
\\&&
z_\Gamma^{\ovl{\rm MS}}\mbox{ (NDR) }
= V_\Gamma^{\ovl{\rm MS}}+1/2 = 0\mbox{ (VA) }, 
5/2\mbox{ (SP) }, 1/2\mbox{ (T) },
\\&&
z_\Gamma^{\rm lat} = V_\Gamma + \Sigma_1
 = \dfrac{h_2(\Gamma)}{4}\log \pi^2 + 16\pi^2 I_\Gamma +\Sigma_1,
\\&&
I_\Gamma = \int_k \left\{
\dfrac{4}{(1-w_0e^{-\alpha})^2 G_0^2}
\left[ e^{-2\alpha} T_3 - 2 \dfrac{e^{-\alpha}}{F_0} T_1
+ \dfrac{X_\Gamma}{F_0^2}\right]
- \dfrac{h_2(\Gamma)}{4}\dfrac{\theta(\pi^2-k^2)}{(k^2)^2}\right\},
\\&&
X_\Gamma = T_4 s^2\mbox{ (SP) },\quad  T_2 \mbox{ (VA) }, \quad 
\dfrac{4T_2-T_4 s^2}{3} \mbox{ (T) }.
\end{eqnarray}

\subsection{4 quark operators (DRED)}

We consider the following $\Delta S=2$ four-quark operators,
\begin{eqnarray}
O_\pm & = & \frac{1}{2} \left[
(\bar q_1 \gamma_\mu^L q_2)(\bar q_3 \gamma_\mu^L q_4)
\pm
(\bar q_1 \gamma_\mu^L q_4)(\bar q_3 \gamma_\mu^L q_2) \right], \\
O_1 & = & 
-C_F (\bar q_1 \gamma_\mu^L q_2)(\bar q_3 \gamma_\mu^R q_4)
+ (\bar q_1 T^a \gamma_\mu^L q_2)(\bar q_3 T^a \gamma_\mu^R q_4), \\
O_2 & = & 
\frac{1}{2N} (\bar q_1 \gamma_\mu^L q_2)(\bar q_3 \gamma_\mu^R q_4)
+ (\bar q_1 T^a \gamma_\mu^L q_2)(\bar q_3 T^a \gamma_\mu^R q_4) ,
\end{eqnarray}
where $\gamma_\mu^{L,R} = \gamma_\mu P_{L,R}$.
The operator matching relation between the lattice and the continuum in
$\msbar$ scheme with DRED is given by
\begin{eqnarray}
O_{4\Gamma}^{\ovl{\rm MS}}(\mu) &=&
(1-w_0^2)^{-2}Z_w^{-2} u^2 Z_{4\Gamma}(\mu a) O_{4\Gamma}^{\rm lat}(1/a),
\end{eqnarray}
where the renormalization factor is given by
\begin{eqnarray}
Z_{4\Gamma} &=& 1+ \dfrac{g^2}{16\pi^2}\left[
(\delta_\Gamma - 2 C_F)\log (\mu a)^2 + z_{4\Gamma}^{\rm MF}
\right],
\\
z_{4\Gamma}^{\rm MF} &=& v_\Gamma^{\ovl{\rm MS}}-v_\Gamma
 + 2 C_F (\Sigma_1^{\ovl{\rm MS}} -\Sigma_1) + 16\pi^2 C_F T_{\rm MF},
\label{eqn:z4}
\\
v_+ &=& \dfrac{N-1}{N}\left[ (N+2) V_{VA}- V_{SP}\right], 
\quad v_+^{\ovl{\rm MS}}= \dfrac{(N-2)(N-1)}{N},
\\
v_- &=& \dfrac{N+1}{N}\left[ (N-2) V_{VA}+ V_{SP}\right],
\quad v_-^{\ovl{\rm MS}}= \dfrac{(N+2)(N+1)}{N},
\\
v_1 &=& N V_{VA}-\dfrac{ V_{SP}}{N},
\quad v_1^{\ovl{\rm MS}}= \dfrac{(N-2)(N+2)}{N},
\\
v_2 &=& \dfrac{N^2-1}{N} V_{SP},
\quad v_2^{\ovl{\rm MS}}= \dfrac{4(N-1)(N+1)}{N},
\\
\delta_\Gamma &=& v_\Gamma^{\msbar} .
\end{eqnarray}
Here we notice that the finite part on lattice can be written in terms
of the one loop correction to bilinear operators.
The relation to the NDR is given in the following subsection.

\subsection{3 quark operators (DRED)}

We consider the three-quark operators relevant to the proton decay
amplitude,
\begin{eqnarray}
\left(O_{PD}\right)_\delta =\varepsilon^{abc} 
\left((\bar q^c_1)^a \Gamma_X (q_2)^b\right) (\Gamma_Y (q_3)^c)_\delta
\label{eq:O_PD}
\end{eqnarray}
where $\bar q^c = -q^T C^{-1}$ with $C=\gamma_0\gamma_2$ is a
charge conjugated field of $q$ and 
$\Gamma_X\otimes\Gamma_Y = P_R\otimes P_R, P_R\otimes P_L, 
P_L\otimes P_R, P_L\otimes P_L$. 
$\otimes$ acts on the Dirac spinor space representing 
$[\gamma_X \otimes \gamma_Y ]_{\alpha\beta;\gamma\delta} \equiv
(\gamma_X)_{\alpha\beta}(\gamma_Y)_{\gamma\delta}$. 

The operator matching relation for the three-quark operators is given as 
follows in $\msbar$ scheme with DRED.
\begin{eqnarray}
O_{PD}^{\ovl{\rm MS}}(\mu) &=& (1-w_0^2)^{-3/2}Z_w^{-3/2}
 u^{3/2} Z_{PD}(\mu a)
O_{PD}^{\rm lat}(1/a),
\end{eqnarray}
where the renormalization factor is
\begin{eqnarray}
Z_{PD} &=& 1+ \dfrac{g^2}{16\pi^2}\left[
(\delta_{PD} - \dfrac{3}{2} C_F)\log (\mu a)^2 + z_{PD}^{\rm MF}
\right],
\\
z_{PD}^{\rm MF} &=& v_{PD}^{\ovl{\rm MS}}-v_{PD} + \dfrac{3}{2}
C_F (\Sigma_1^{\ovl{\rm MS}}-\Sigma_1) + 16\pi^2 C_F \dfrac{3T_{\rm MF}}{4},
\\
v_{PD} &=& \dfrac{N+1}{2N}\left[ 2 V_{VA}+ V_{SP}\right],
\quad v_{PD}^{\ovl{\rm MS}}= \delta_{PD}=6\dfrac{N+1}{2N}.
\end{eqnarray}

\subsection{Renormalization factor for $B_K$ and $B_P$}

Following quantities are important for $K$ mesons.
One is $K$ meson B parameter $B_K$, defined by
\begin{eqnarray}
B_K =
\frac{
\VEV{K}{\ovl{s}\gamma_\mu(1-\gamma_5)d \ovl{s}\gamma_\mu(1-\gamma_5)d}{K}}
{\frac{8}{3}\VEV{K}{\ovl{s}\gamma_\mu\gamma_5d}{0}
 \VEV{0}{\ovl{s}\gamma_\mu\gamma_5d}{K}},
\label{eqn:BK}
\end{eqnarray}
which is needed to extract the CKM matrix from experiments,
and the other is the matrix element divided by the pseudo scalar density, 
\begin{eqnarray}
B_P =
\frac{
\VEV{K}{\ovl{s}\gamma_\mu(1-\gamma_5)d \ovl{s}\gamma_\mu(1-\gamma_5)d}{K}}
{\VEV{K}{\ovl{s}\gamma_5d}{0} \VEV{0}{\ovl{s}\gamma_5d}{K}},
\label{eqn:BP}
\end{eqnarray}
which can be used to measure the violation of the chiral symmetry,
since it should vanish at $m_\pi\to 0$ in the presence of the chiral symmetry.
The $s$ and $d$ quark fields defining these quantities are the
boundary fields given by (\ref{eq:quark}) and these parameters are
written in terms of the four-quark operator $O_+$ and bilinear
quark operators $O_A$, $O_P$ in the previous subsection.
The renormalization factors for $B_K$ and $B_P$ are given by the ratio
of those for $O_+$ and $O_A$, $O_P$.
\begin{eqnarray}
Z_{B_K}( \mu a ) &=&\frac{(1-w_0^2)^{-2} Z_w^{-2} Z_+ (\mu a )}
{(1-w_0)^{-2} Z_w^{-2} Z_A(\mu a)^2}
=\frac{Z_+ (\mu a )}{Z_A(\mu a)^2}
\nn\\&=&
1+\frac{g^2}{16\pi^2}\left[ -4\log (\mu a) + z_{B_K}^{\rm MF}\right],
\\
Z_{B_P}( \mu a ) &=&\frac{(1-w_0^2)^{-2} Z_w^{-2} Z_+ (\mu a )}
{(1-w_0)^{-2} Z_w^{-2} Z_P(\mu a)^2}
=\frac{Z_+ (\mu a )}{Z_P(\mu a)^2}
\nn\\&=&
1 +\frac{g^2}{16\pi^2}\left[ -20\log (\mu a)+z_{B_P}^{\rm MF} \right],
\end{eqnarray}
where
\begin{eqnarray}
&&
z_{B_K}^{\rm MF} = z_+^{\rm MF} -2 C_F z_A^{\rm MF},
\\&&
z_{B_P}^{\rm MF} = z_+^{\rm MF} -2 C_F z_P^{\rm MF}.
\end{eqnarray}
From the explicit notation of \eqn{eqn:zb} and \eqn{eqn:z4} the mean field
improvement factor $T_{\rm MF}$ is canceled out in $z_{B_K}^{\rm MF}$ and
$z_{B_P}^{\rm MF}$.
The remaining effect of the mean field improvement is 
to shift the domain-wall height $M$.

\subsection{DRED and NDR}

In this subsection we list the relation between NDR and DRED in
$\ovl{\rm MS}$ scheme.
The relations for the quark wave function and the quark mass are given by
\begin{eqnarray}
z_2({\rm NDR}) &=& z_2({\rm DRED}) +1,
\\
z_m({\rm NDR}) &=& z_m({\rm DRED}) +1.
\end{eqnarray}
The relations for the bilinear quark operators are
\begin{eqnarray}
z_{\rm VA}({\rm NDR}) &=& z_{\rm VA}({\rm DRED}) - 1/2,
\\
z_{\rm SP}({\rm NDR}) &=& z_{\rm SP}({\rm DRED}) - 1,
\\
z_{\rm T}({\rm NDR}) &=& z_{\rm T}({\rm DRED}) + 1.
\end{eqnarray}
The renormalization factors for the four-quark operators are related
with
\begin{eqnarray}
z_+({\rm NDR}) &=& z_+({\rm DRED}) -3,
\\
z_-({\rm NDR}) &=& z_-({\rm DRED}) +2,
\\
z_{ij}({\rm NDR}) &=& 
\left(
\begin{array}{cc}
z_1({\rm DRED})  & 0 \\
0   & z_2({\rm DRED})  \\
\end{array}
\right)
+
\left(
\begin{array}{cc}
5/6  & 8 \\
1   & -8/3  \\
\end{array}
\right).
\end{eqnarray}
The relation for the three-quark operator is
\begin{eqnarray}
z_{\rm PD}({\rm NDR}) &=& z_{\rm PD}({\rm DRED}) + 2/3.
\end{eqnarray}
At last we have a relation for $B_K$ and $B_P$
\begin{eqnarray}
&&
z_{B_K}({\rm NDR}) = z_{B_K}({\rm DRED}) -5/3,
\\&&
z_{B_P}({\rm NDR}) = z_{B_P}({\rm DRED}) -1/3.
\end{eqnarray}
Here we take the color factor $N=3$ for three- and four-quark
operators and $B_K$, $B_P$.

\section{Numerical results}
\label{sec:numerical}

Finite parts of renormalization factors in the previous section
are numerically calculated. Necessary momentum integration is approximated
by discrete sum of $L^4$  with $L=64$.
For three- and four-quark operators and the parameters $B_K$, $B_P$ the
color factor is set to $N=3$.
In the tables\ \ref{tab:Plaq}$ \sim$ \ref{tab:BK2},
numerical values in DRED schemes without mean field (MF)
improvement is given.
The numerical error is estimated by varying $L$ from $64$ to $60$.

The renormalization factors with the MF improvement defined in the previous
section are given by shift in the domain-wall height and subtraction of
the tadpole factor $T_{\rm MF}$ as follows.
\begin{eqnarray}
\Sigma_w^{\rm MF} &= & \Sigma_w|_{w_0=w_0^{\rm MF}} + (16\pi^2) 2 T_{\rm MF},
\\
z_2^{\rm MF} &=& z_2|_{w_0=w_0^{\rm MF}} + 16\pi^2 \frac{1}{2}T_{\rm MF},
\\
z_m^{\rm MF} &=& z_m|_{w_0=w_0^{\rm MF}} - 16\pi^2 \frac{1}{2}T_{\rm MF},
\\
z_\Gamma^{\rm MF} &=& z_\Gamma|_{w_0=w_0^{\rm MF}}
 + 16\pi^2 \frac{1}{2}T_{\rm MF},
\quad \Gamma =\mbox{ VA, SP, T },
\\
z_{4\Gamma}^{\rm MF} &=& z_{4\Gamma}|_{w_0=w_0^{\rm MF}}
 + 16\pi^2 C_F T_{\rm MF},
\quad \Gamma = \pm, 1, 2,
 \\
z_{\rm PD}^{\rm MF} &=& z_{\rm PD}|_{w_0=w_0^{\rm MF}}
 + 16\pi^2 C_F \frac{3}{4} T_{\rm MF},
\end{eqnarray}
where $w_0=1-M$ and a nonperturbatively shifted $w_0^{\rm MF}$ 
is defined by
\begin{eqnarray}
 w_0^{\rm MF}=1-w_0+4(1-u).
\end{eqnarray}

\section{Tool kit for the mean field improved perturbation theory}
\label{sec:tool}

In this section, we explain step by step 
how to calculate the renormalization factors in the mean field improved 
perturbation theory, using our numerical results in the previous section.
We assume the (quenched) domain-wall QCD with the plaquette or 
improved gauge actions.
Therefore we have two parameters, the bare gauge coupling constant
$g^2 = 6/\beta$ and the domain-wall height $M$, in addition to the
improved coefficient $c_1$ and $c_2+c_3$.

\begin{enumerate}
\item First of all, one determines the tadpole factor $u$. Usually
$u$ is determined from the measured plaquette value such that
$ u = P^{1/4}$. 
From the existing data for the plaquette values at several $\beta$
for the plaquette and RG improved (Iwasaki) gauge actions,
$P$ may be parameterized as the function of $g^2$:
\beqa
 P &=& \dfrac{ 1+a_1 g^2 + a_2 g^4 + a_3 g^6 + a_4 g^8}{1+ b_1 g^2}.
\label{eq:interP}
\eeqa
In table\ref{tab:PMF} the coefficient $a_i$ and $b_1$ are given\cite{kaneko},
where $g^2_{\rm max}$ is the maximum value of the coupling used to the fit.
The fits are constrained to satisfy the relation that $ c_p = b_1-a_1$,
where 
\beqa
P = 1 - c_p g^2 +O(g^4)
\eeqa
at the leading order of the perturbation theory.
The value of $c_p$ for various gauge actions is given 
in table~\ref{tab:pert}, together with
the value of $c_{R1,R3}$, defined by
\beqa
R1 &=& \frac{1}{3}{\rm Tr} U_{rtg}=1 - c_{R1} g^2 +O(g^4)\\
R2 &=& \frac{1}{3}{\rm Tr} U_{chr}=1 - c_{R2} g^2 +O(g^4)\\
R3 &=& \frac{1}{3}{\rm Tr} U_{plg}=1 - c_{R3} g^2 +O(g^4)
\eeqa
where $R1$, $R2$ and $R3$ are the expectation value of the rectangular,
chair and parallelogram loops, respectively.

At the strong coupling such that $g^2 > g^2_{\rm max}$,
one had better to use the measured value of $P$, instead of 
eq.~(\ref{eq:interP}).

\item One then calculates the mean-field improved $\overline{\rm MS}$ 
coupling $g_{\overline{\rm MS}}^2(\mu )$ at the scale $\mu$
from $g^2$ and $P$ as
\beqa
\dfrac{1}{g_{\overline{\rm MS}}^2(\mu )}
&=& \dfrac{P}{g^2} + d_g + c_p +\dfrac{11}{8\pi^2} \log (\mu a)
\eeqa
where 
$d_g$, taken from Refs.\cite{IS,IY,SSN}, is also given in table~\ref{tab:pert}.
According to the philosophy of the mean-field improvement,
an alternative formulae seems more natural for the RG improved gauge 
action\cite{cppacs}:
\beqa
\dfrac{1}{g_{\overline{\rm MS}}^2(\mu )}
&=& \dfrac{c_0 P + 8 c_1 R1+ 16c_2 R2 +8 c_3 R3}{g^2} \nn \\
& &+ d_g + (c_0\cdot c_p + 8 c_1\cdot c_{R1}+16 c_2\cdot c_{R2}+
8 c_3\cdot c_{R3})
+\dfrac{11}{8\pi^2} \log (\mu a) .
\eeqa

\item One replaces the domain-wall height $M$ with $\tilde M =
M - 4(1-u)$.

\item Wave function renormalization factors are 
given by
\beqa
q_{\msbar} & = & [(1-w_0^2)Z_w]^{-1/2} ( u Z_2)^{1/2} q_{\rm lat},\\
Z_w &=& 1+ \dfrac{g_{\msbar}^2}{16\pi^2} C_F z_w^{\rm MF}, \\
z_w^{\rm MF} &=& \dfrac{2 w_0}{1-w_0^2} 
\left[\Sigma_w + 32\pi^2 T_{MF}\right], \\
Z_2 &=& 1 + \dfrac{g_{\msbar}^2}{16\pi^2} C_F
\left[-2 \log (\mu a) + z_2^{\rm MF}\right],\\
z_2^{\rm MF} &=&\Sigma_1^{\msbar}-\Sigma_2 + 16\pi^2 \dfrac{T_{\rm MF}}{2},
\eeqa
where $w_0 = 1 - \tilde M$. We parameterize $\Sigma_w+ 32\pi^2 T_{\rm MF}$ and
$z_2^{\rm MF}$ as follows:
\beqa
\Sigma_w +32\pi^2 T_{\rm MF} &=& 32\pi^2 T_{\rm MF} + a_0 
+ a_1 w_0 + a_2  w_0^2 +  a_3  w_0^3 +  a_4  w_0^4 
=32\pi^2 T_{\rm MF} +\sum_{i=0}^4 a_i w_0^i, \\
z_2^{\rm MF} &=& 8\pi^2 T_{\rm MF} +\sum_{i=0}^{4,\ 6} a_i w_0^i.
\eeqa
The relative errors of these and the following interpolations 
are less than a few \% except at points where the value is near zero.
We have collected the values of parameters $a_i$ for
$\Sigma_w$ in table~\ref{tab:inter_zw} and those for $z_2^{\rm MF}$
in table~\ref{tab:inter_z2}. Note that $z_2^{\rm MF}$ depends on
the detail of the $\msbar$ scheme, NDR and DRED, hence both results
of $a_0$ are given.

\item The mass renormalization factor is given by
\beqa
m_{\msbar} &=& (1-w_0^2)Z_w Z_m \dfrac{m_{\rm lat}}{u} \\
Z_m &=& 1 + \dfrac{g_{\msbar}^2}{16\pi^2} C_F 
\left[ -6\log (\mu a) + z_m^{\rm MF}\right]\\
z_m^{\rm MF} &=& (\Sigma_2^{\msbar}-\Sigma_2)-(\Sigma_1^{\msbar}-\Sigma_1)
-16\pi^2\dfrac{T_{\rm MF}}{2} = -8\pi^2 T_{\rm MF} +\sum_{i=0}^{4,\ 6,\ 8} a_i w_0^i.
\eeqa
The values of $a_i$ are given in table~\ref{tab:inter_zm}.

\item The renormalization factors for the bilinear operators are given by
\beqa
O_{\Gamma}^{\msbar} &=& \dfrac{u}{(1-w_0^2)Z_w}Z_\Gamma (\mu a)
O_{\Gamma}^{\rm lat}, \\
Z_\Gamma &=& 1 +\dfrac{g_\msbar^2}{16\pi^2}C_F
\left[h(\Gamma)\log (\mu a) + z_{\Gamma}^{\rm MF}\right], \\
z_\Gamma^{\rm MF} &=& z_\Gamma^\msbar - z_\Gamma + 16\pi^2\dfrac{T_{\rm MF}}{2}
= 8\pi^2 T_{\rm MF} +\sum_{i=0}^{4,\ 6,\ 8} a_i w_0^i 
\eeqa
where $h(\Gamma)$ = 0 (VA), 6(SP), -2(T).
The values of $a_i$ are given in table~\ref{tab:inter_zbi}.

\item The renormalization factors for the 4-quark operators
are given by
\beqa
O^{\msbar}_I &=& \dfrac{u^2}{[(1-w_0^2)Z_w]^2}
Z_{IJ}^{4\Gamma}(\mu a)O_J^{\rm lat}\\
Z_{IJ}^{4\Gamma}&=& \delta_{IJ}\left\{1 +\dfrac{g_\msbar^2}{16\pi^2}
\left[ \delta_I \log(\mu a) +z_{I}^{\rm MF}\right]\right\}
+\dfrac{g_\msbar^2}{16\pi^2} v_{IJ} \\
z_{I}^{\rm MF}&=&v_I^\msbar-v_I + 2 C_F(\Sigma_1^\msbar-\Sigma_1)+
16\pi^2 C_F T_{\rm MF} \nn \\
&=& 16\pi^2 C_F T_{\rm MF} +\sum_{i=0}^{4,\ 8} a_i w_0^i 
\eeqa
where $I=+,-,1,2$,
$\delta_+ =-4$, $\delta_- = 8$, $\delta_1 = -2$, $\delta_2 =16$,
and $ v_{IJ} = 0$ for all $I$ and $J$ except
$v_{12} = 8$ and $v_{21}=1$ in the NDR scheme.
The values of $a_i$ are given in table~\ref{tab:inter_z4q}.
Note that the interpolation of $z_2$ by the 4th polynomial of $w_0$ is 
not accurate enough for the DBW2 gauge action, so we also employ 
the 6th polynomial of $w_0$.

\item The renormalization factor for the 3 quark operator
is given by
\beqa
O^{\msbar}_{PD} &=& \dfrac{u^{3/2}}{[(1-w_0^2)Z_w]^{3/2}}
Z_{PD}(\mu a)O_{PD}^{\rm lat}\\
Z_{PD}&=& 1 +\dfrac{g_\msbar^2}{16\pi^2}
\left[ 4 \log(\mu a) +z_{PD}^{\rm MF}\right] \\
z_{PD}^{\rm MF}&=&v_{PD}^\msbar-v_{PD} + \dfrac{3}{2}C_F
(\Sigma_1^\msbar-\Sigma_1)+16\pi^2 C_F \dfrac{3}{4}T_{\rm MF} 
= 12\pi^2 C_F T_{\rm MF} +\sum_{i=0}^{4,\ 6} a_i w_0^i .
\eeqa
The values of $a_i$ are given in table~\ref{tab:inter_z3q}.

\item The renormalization factors for $B_K$ and $B_P$ are given by
\beqa
Z_{B_K}(\mu a) &=& \dfrac{Z_+(\mu a)}{Z_A(\mu a)^2}
= 1 + \dfrac{g_\msbar^2}{16\pi^2}\left[-4\log(\mu a) + z_{B_K}^{\rm MF}\right],\\
z_{B_K}^{\rm MF} &=& z_+^{\rm MF} - 2 C_F z_A^{\rm MF}
= \sum_{i=0}^{9} a_i w_0^i , \\
Z_{B_P}(\mu a) &=& \dfrac{Z_+(\mu a)}{Z_P(\mu a)^2}
= 1 + \dfrac{g_\msbar^2}{16\pi^2}\left[-20\log(\mu a) + z_{B_P}^{\rm MF}\right],\\
z_{B_P}^{\rm MF} &=& z_+^{\rm MF} - 2 C_F z_P^{\rm MF}
= \sum_{i=0}^{9} a_i w_0^i . \\
\eeqa
The values of $a_i$ are given in table~\ref{tab:inter_zbk}.

\item The renormalization factors for 4-quark operators with $\Delta S =1$ 
are defined as
\beqa
Q_i^{\msbar} &=& \dfrac{1}{[(1-w_0) Z_w]^2}
\left[ Z_{ij}^g Q_j^{\rm lat} + Z_i^{\rm pen}Q_{\rm pen}^{\rm lat}\right]
\eeqa
where $i,j = 1,2,\cdots , 10$.
The renormalization factor for the penguin operators is given 
by\cite{aoki-kuramashi}
\beqa
Z_i^{\rm pen} &=& \dfrac{g_\msbar^2}{48\pi^2}C(Q_i)
\left[ -2\log (\mu a)^2 + z_i^{\rm pen}\right]\\
Q_{\rm pen}^{\rm lat}&=& Q_4 + Q_6 -\dfrac{1}{N}(Q_3+Q_5) \\
z_i^{\rm pen} &=&\frac{\sum_{i=0}^4 a_i w_0^i}{1+\sum_{i=1}^4 b_i w_0^i} ,
\eeqa
where $C(Q_2)=1$, $C(Q_3)=2$, $C(Q_4)=C(Q_6)=3$, $C(Q_9)=-1$
and $C(Q_i)=0$ for other $i$.
The renormalization factor $z_i^{\rm pen}$ is independent of the gauge 
action. Furthermore it does not depend on $i$ for DRED, but
\beqa
z_{4,6,8,10}^{\rm pen}({\rm NDR}) &=& z_i^{\rm pen}({\rm DRED}) +1/4\\
z_{1,2,3,5,7,9}^{\rm pen}({\rm NDR}) &=& z_i^{\rm pen}({\rm DRED}) +5/4 .
\eeqa
Numerical values for $z_i^{\rm pen}$(DRED) are given in table~\ref{tab:pen},
together with interpolated values and relative errors defined by
($z_i^{\rm pen}$ - interpolation)/$z_i^{\rm pen}$.
The values of $a_i$ and $b_i$ for the interpolation are also given in table~\ref{tab:pen}.

For other renormalization factors, we have
\beqa
Z_{ii}^g &=& u^2 \times \left\{
\begin{array}{ll}
1+\dfrac{1}{16\pi^2}g^2_\msbar\left[
2\log(\mu a) + z_{11}^{\rm MF}\right]
&\quad i=1,2,3,4,9,10, \\
1+\dfrac{1}{16\pi^2}g^2_\msbar\left[
-2\log(\mu a) + z_{55}^{\rm MF}\right]
&\quad i=5,7, \\
1+\dfrac{1}{16\pi^2}g^2_\msbar\left[
16\log(\mu a) +z_{66}^{\rm MF}\right]
&\quad i=6,8, \\
\end{array}\right.
\eeqa
\beqa
Z_{ij}^g &=& \left\{
\begin{array}{ll}
= Z_{ji}^g=\dfrac{1}{16\pi^2}g^2_\msbar
\left[-6 \log(\mu a) +z_{12}^{\rm MF}\right]
&\quad (i,j) =(1,2),(3,4),(9,10), \\
\dfrac{1}{16\pi^2}g^2_\msbar
\left[6 \log(\mu a) +z_{56}^{\rm MF}\right]
&\quad (i,j) =(5,6),(7,8), \\
\dfrac{1}{16\pi^2}g^2_\msbar z_{65}^{\rm MF}
&\quad (i,j) =(6,5),(8,7), \\
0 &\quad\mbox{others} \\
\end{array}\right.
\eeqa
where
\beqa
z_{11}^{\rm MF}&=&\dfrac{z_+^{\rm MF} + z_-^{\rm MF}}{2} = 
16\pi^2C_F T_{\rm MF}+\sum_{i=0}^{4,\ 6} a_i w_0^i, \\
z_{55}^{\rm MF}&=&z_1^{\rm MF}-v_{21} = 
16\pi^2C_F T_{\rm MF}+\sum_{i=0}^4 a_i w_0^i, \\
z_{66}^{\rm MF}&=&z_2^{\rm MF}+v_{21} = 
16\pi^2C_F T_{\rm MF}+\sum_{i=0}^{4,\ 8} a_i w_0^i, \\
z_{12}^{\rm MF}&=&\dfrac{z_+^{\rm MF} - z_-^{\rm MF}}{2} 
= \sum_{i=0}^{9} a_i w_0^i,\\
z_{56}^{\rm MF}&=&\dfrac{z_2^{\rm MF}-z_1^{\rm MF}+v_{21}-v_{12}}{3}
\\
z_{65}^{\rm MF}&=& -3 v_{21} .
\eeqa
The values of $a_i$ are given in table~\ref{tab:delS}.
Note that 
\beqa
z_{56}^{\rm MF}({\rm DRED}) &=& - z_{12}^{\rm MF}({\rm DRED}) ,\\
v_{12}({\rm DRED}) &=& v_{21}({\rm DRED}) = 0 ,
\eeqa
and that
\beqa
z_{11}({\rm NDR}) &=& z_{11}({\rm DRED}) -\frac{1}{2} ,\\
z_{55}({\rm NDR}) &=& z_{55}({\rm DRED}) -\frac{1}{6} ,\\
z_{66}({\rm NDR}) &=& z_{66}({\rm DRED}) -\frac{5}{3} ,\\
z_{12}({\rm NDR}) &=& z_{12}({\rm DRED}) -\frac{5}{2} ,\\
z_{56}({\rm NDR}) &=& z_{56}({\rm DRED}) -\frac{7}{2}= -z_{12}({\rm DRED}) -\frac{7}{2} ,\\
z_{65}({\rm NDR}) &=& z_{65}({\rm DRED}) -3 = -3 .
\eeqa

\end{enumerate}

If one uses the measured averaged link variable for $u$ 
instead of $P= u^{1/4}$, 
one should replace $T_{\rm MF}$ in the above formula with
\beqa
T_{\rm MF}&\rightarrow& T-(1-\alpha)\delta T_{\rm gauge}
\eeqa
where $\alpha$ is the gauge parameter, $\delta T_{\rm gauge}
=0.0387334$ for all gauge actions, and $T$ is given in
table~\ref{tab:pert}.
Similarly, in the case of using the critical hopping parameter 
$K_c$ of the Wilson fermion for $u$, one should make the replacement that
$T_{\rm MF} \rightarrow T_{K_c}$ ,
where $T_{K_c}-T_{\rm MF}$  is given also in table~\ref{tab:pert}.

If one set $u=1$ and $T_{\rm MF}=0$ in the above steps,
one can obtain the renormalization factors
in the ordinary perturbation theory without MF improvement.

\section{Conclusion}
\label{sec:conclusion}

In this paper we have evaluated renormalization factors at the one loop level
for quark wave function, quark mass, bilinear quark operators, four- and
three-quark operators perturbatively.
We show that the mean field improvement, by which
the large additive quantum correction to the domain-wall height $M$
is non-perturbatively estimated,  makes the
perturbative evaluations more reliable.
We explain how to obtain the numerical values for the renormalization factors 
in the MF improved perturbation theory in detail.

\section*{Acknowledgments}

This work is supported in part by Grants-in-Aid
of the Ministry of Education (Nos.12640253,13135202,14046202).
T.I wants to thank Dr. Chris Dawson for illuminating discussions.

\newcommand{\J}[4]{{#1} {#2} #3 (#4)}
\newcommand{\AP}{Ann.~Phys.}
\newcommand{\CMP}{Commun.~Math.~Phys.}
\newcommand{\IJMP}{Int.~J.~Mod.~Phys.}
\newcommand{\MPL}{Mod.~Phys.~Lett.}
\newcommand{\NP}{Nucl.~Phys.}
\newcommand{\NPSup}{Nucl.~Phys.~B (Proc.~Suppl.)}
\newcommand{\PL}{Phys.~Lett.}
\newcommand{\PR}{Phys.~Rev.}
\newcommand{\PRL}{Phys.~Rev.~Lett.}
\newcommand{\PTP}{Prog. Theor. Phys.}
\newcommand{\Suppl}{Prog. Theor. Phys. Suppl.}

\tightenlines
\begin{table}
\caption{Results for Plaquette ($c_1=c_2=c_3=0$) without mean filed
 improvement.}  
\label{tab:Plaq}
\begin{center}
\begin{tabular}{|l|ll|l|lll|}
$M$ & $z_2$(DRED) & $z_m$(DRED) & $\Sigma_w$ & 
$z_{V/A}$(DRED) & $z_{S/P}$(DRED) & $z_{T}$(DRED) \\
\hline
$ 0.05$ & $  -13.7495(11)$ & $   13.0515(34)$ & $-51.225797$
 & $-17.582757(74)$ & $  -13.0514(36)$ & $  -19.0932(11)$ \\
$ 0.10$ & $  -13.6613(11)$ & $   13.4795(33)$ & $-51.048195$
 & $-17.4958220(51)$ & $  -13.4795(33)$ & $  -18.8346(11)$ \\
$ 0.15$ & $  -13.5825(11)$ & $   13.8202(33)$ & $-50.889530$
 & $-17.4183180(92)$ & $  -13.8202(33)$ & $  -18.6177(11)$ \\
$ 0.20$ & $  -13.5109(11)$ & $   14.1140(34)$ & $-50.744997$
 & $-17.347925(44)$ & $  -14.1140(35)$ & $  -18.4259(11)$ \\
$ 0.25$ & $  -13.4449(11)$ & $   14.3783(34)$ & $-50.611900$
 & $-17.283312(47)$ & $  -14.3782(35)$ & $  -18.2517(11)$ \\
$ 0.30$ & $  -13.3838(11)$ & $   14.6225(33)$ & $-50.488509$
 & $-17.223679(14)$ & $  -14.6226(32)$ & $  -18.0907(11)$ \\
$ 0.35$ & $  -13.3269(11)$ & $   14.8523(33)$ & $-50.373634$
 & $-17.168268(24)$ & $  -14.8523(34)$ & $  -17.9403(11)$ \\
\end{tabular}
\end{center}
\end{table}
\begin{table}
\caption{Results for Plaquette(continued) without mean filed improvement.}
\label{tab:Plaq_2}
\begin{center}
\begin{tabular}{|l|llll|l|}
 $M$ & $z_{+}$(DRED) & $z_{-}$(DRED) & $z_{1}$(DRED) & $z_{2}$(DRED)
& $z_{PD}$(DRED)\cr
 \hline
$ 0.05$ & $  -49.9083(21)$ & $  -40.8455(49)$ & $ -48.39781(97)$
 & $  -34.8037(96)$ & $  -32.1446(25)$ \\
$ 0.10$ & $  -49.3331(22)$ & $  -41.3005(44)$ & $  -47.9943(11)$
 & $  -35.9454(87)$ & $  -32.3141(22)$ \\
$ 0.15$ & $  -48.8476(22)$ & $  -41.6514(43)$ & $  -47.6482(11)$
 & $  -36.8540(87)$ & $  -32.4379(22)$ \\
$ 0.20$ & $  -48.4171(22)$ & $  -41.9492(47)$ & $  -47.3391(10)$
 & $  -37.6374(92)$ & $  -32.5399(24)$ \\
$ 0.25$ & $  -48.0256(22)$ & $  -42.2154(47)$ & $  -47.0572(10)$
 & $  -38.3419(93)$ & $  -32.6299(24)$ \\
$ 0.30$ & $  -47.6639(22)$ & $  -42.4617(43)$ & $  -46.7968(11)$
 & $  -38.9935(86)$ & $  -32.7133(21)$ \\
$ 0.35$ & $  -47.3260(22)$ & $  -42.6941(46)$ & $  -46.5540(11)$
 & $  -39.6062(90)$ & $  -32.7926(23)$ \\
\end{tabular}
\end{center}
\end{table}

\begin{table}
\caption{Results for Iwasaki ($c_1=-0.331$,  $c_2+c_3=0$) without mean
 filed improvement.} 
\label{tab:RG}
\begin{center}
\begin{tabular}{|l|ll|l|lll|}
$M$ & $z_2$(DRED) & $z_m$(DRED) & $\Sigma_w$ & 
$z_{V/A}$(DRED) & $z_{S/P}$(DRED) & $z_{T}$(DRED) \\
\hline
$ 0.05$ & $   -5.2309(11)$ & $    3.5185(34)$ & $-25.891445$
 & $ -9.041437(74)$ & $   -3.5184(36)$ & $  -10.8825(11)$ \\
$ 0.10$ & $   -5.1519(11)$ & $    3.9209(33)$ & $-25.740748$
 & $-8.9630147(50)$ & $   -3.9210(33)$ & $  -10.6437(11)$ \\
$ 0.15$ & $   -5.0821(11)$ & $    4.2353(33)$ & $-25.608432$
 & $-8.8938444(92)$ & $   -4.2354(33)$ & $  -10.4467(11)$ \\
$ 0.20$ & $   -5.0193(11)$ & $    4.5020(34)$ & $-25.489672$
 & $ -8.831595(44)$ & $   -4.5020(35)$ & $  -10.2748(11)$ \\
$ 0.25$ & $   -4.9619(11)$ & $    4.7383(34)$ & $-25.381743$
 & $ -8.774922(47)$ & $   -4.7382(35)$ & $  -10.1205(11)$ \\
$ 0.30$ & $   -4.9093(11)$ & $    4.9536(33)$ & $-25.282889$
 & $ -8.723012(14)$ & $   -4.9537(32)$ & $   -9.9795(11)$ \\
$ 0.35$ & $   -4.8607(11)$ & $    5.1535(33)$ & $-25.191889$
 & $ -8.675093(24)$ & $   -5.1535(34)$ & $   -9.8490(11)$ \\
\end{tabular}
\end{center}
\end{table}
\begin{table}
\caption{Results for Iwasaki(continued) without mean filed improvement.}
\label{tab:RG_2}
\begin{center}
\begin{tabular}{|l|llll|l|}
 $M$ & $z_{+}$(DRED) & $z_{-}$(DRED) & $z_{1}$(DRED) & $z_{2}$(DRED) &
 $z_{PD}$(DRED)\cr 
\hline
$ 0.05$ & $  -27.7926(21)$ & $  -16.7464(49)$ & $ -25.95153(97)$
 & $   -9.3823(96)$ & $  -14.4008(25)$ \\
$ 0.10$ & $  -27.2627(22)$ & $  -17.1786(44)$ & $  -25.5821(11)$
 & $  -10.4559(87)$ & $  -14.5647(22)$ \\
$ 0.15$ & $  -26.8226(22)$ & $  -17.5056(43)$ & $  -25.2697(11)$
 & $  -11.2943(87)$ & $  -14.6820(22)$ \\
$ 0.20$ & $  -26.4373(22)$ & $  -17.7781(47)$ & $  -24.9941(10)$
 & $  -12.0053(92)$ & $  -14.7768(24)$ \\
$ 0.25$ & $  -26.0909(22)$ & $  -18.0175(47)$ & $  -24.7454(10)$
 & $  -12.6352(93)$ & $  -14.8587(24)$ \\
$ 0.30$ & $  -25.7743(22)$ & $  -18.2356(43)$ & $  -24.5178(11)$
 & $  -13.2097(86)$ & $  -14.9331(21)$ \\
$ 0.35$ & $  -25.4813(22)$ & $  -18.4381(46)$ & $  -24.3074(11)$
 & $  -13.7427(90)$ & $  -15.0025(23)$ \\
\end{tabular}
\end{center}
\end{table}

\begin{table}
\caption{Results for DBW2 ($c_1=-1.407$,  $c_2+c_3=0$) without mean
 filed improvement.}
\label{tab:DBW2}
\begin{center}
\begin{tabular}{|l|ll|l|lll|}
$M$ & $z_2$(DRED) & $z_m$(DRED) & $\Sigma_w$ & 
$z_{V/A}$(DRED) & $z_{S/P}$(DRED) & $z_{T}$(DRED) \\
\hline
$ 0.05$ & $   -0.0488(11)$ & $   -3.1867(34)$ & $-11.378295$
 & $ -3.847303(77)$ & $    3.1868(36)$ & $   -6.1920(11)$ \\
$ 0.10$ & $    0.0152(11)$ & $   -2.8238(33)$ & $-11.263725$
 & $-3.78359680(17)$ & $    2.8237(33)$ & $   -5.9860(11)$ \\
$ 0.15$ & $    0.0704(11)$ & $   -2.5490(33)$ & $-11.166177$
 & $-3.7286926(92)$ & $    2.5490(33)$ & $   -5.8212(11)$ \\
$ 0.20$ & $    0.1190(11)$ & $   -2.3223(33)$ & $-11.080847$
 & $ -3.680251(25)$ & $    2.3223(34)$ & $   -5.6811(11)$ \\
$ 0.25$ & $    0.1626(11)$ & $   -2.1264(33)$ & $-11.005031$
 & $ -3.636919(24)$ & $    2.1264(34)$ & $   -5.5580(11)$ \\
$ 0.30$ & $    0.2020(11)$ & $   -1.9518(33)$ & $-10.936985$
 & $ -3.597872(14)$ & $    1.9518(32)$ & $   -5.4478(11)$ \\
$ 0.35$ & $    0.2378(11)$ & $   -1.7932(33)$ & $-10.875495$
 & $ -3.562327(24)$ & $    1.7932(34)$ & $   -5.3475(11)$ \\
\end{tabular}
\end{center}
\end{table}
\begin{table}
\caption{Results for DBW2(continued) without mean filed improvement.}
\label{tab:DBW2_2}
\begin{center}
\begin{tabular}{|l|llll|l|}
 $M$ & $z_{+}$(DRED) & $z_{-}$(DRED) & $z_{1}$(DRED) & $z_{2}$(DRED) &
 $z_{PD}$(DRED)\cr 
\hline
$ 0.05$ & $  -14.9489(21)$ & $   -0.8806(49)$ & $ -12.60419(97)$
 & $    8.4982(96)$ & $   -3.0052(25)$ \\
$ 0.10$ & $  -14.4945(22)$ & $   -1.2798(44)$ & $  -12.2920(11)$
 & $    7.5299(88)$ & $   -3.1623(22)$ \\
$ 0.15$ & $  -14.1283(22)$ & $   -1.5730(43)$ & $  -12.0357(11)$
 & $    6.7973(87)$ & $   -3.2723(22)$ \\
$ 0.20$ & $  -13.8157(22)$ & $   -1.8106(46)$ & $  -11.8149(11)$
 & $    6.1928(91)$ & $   -3.3588(23)$ \\
$ 0.25$ & $  -13.5407(22)$ & $   -2.0140(46)$ & $  -11.6196(11)$
 & $    5.6705(90)$ & $   -3.4316(23)$ \\
$ 0.30$ & $  -13.2941(22)$ & $   -2.1948(43)$ & $  -11.4442(11)$
 & $    5.2047(86)$ & $   -3.4960(21)$ \\
$ 0.35$ & $  -13.0699(22)$ & $   -2.3588(46)$ & $  -11.2847(11)$
 & $    4.7820(90)$ & $   -3.5543(23)$ \\
\end{tabular}
\end{center}
\end{table}

\begin{table}
\caption{Results for Symanzik ($c_1=-1/12$, $c_2+c_3=0$) without mean
 filed improvement.} 
\label{tab:Sy}
\begin{center}
\begin{tabular}{|l|ll|l|lll|}
$M$ & $z_2$(DRED) & $z_m$(DRED) & $\Sigma_w$ & 
$z_{V/A}$(DRED) & $z_{S/P}$(DRED) & $z_{T}$(DRED) \\
\hline
$ 0.05$ & $  -10.1312(11)$ & $    9.1044(33)$ & $-40.243366$
 & $-13.9545650(79)$ & $   -9.1044(33)$ & $  -15.5713(11)$ \\
$ 0.10$ & $  -10.0458(11)$ & $    9.5240(33)$ & $-40.074665$
 & $-13.8700910(49)$ & $   -9.5240(33)$ & $  -15.3188(11)$ \\
$ 0.15$ & $   -9.9698(11)$ & $    9.8563(33)$ & $-39.924749$
 & $-13.7950750(89)$ & $   -9.8563(33)$ & $  -15.1080(11)$ \\
$ 0.20$ & $   -9.9009(11)$ & $   10.1413(33)$ & $-39.788806$
 & $-13.727114(24)$ & $  -10.1412(34)$ & $  -14.9224(11)$ \\
$ 0.25$ & $   -9.8376(11)$ & $   10.3964(33)$ & $-39.664130$
 & $-13.664880(14)$ & $  -10.3964(33)$ & $  -14.7544(11)$ \\
$ 0.30$ & $   -9.7791(11)$ & $   10.6311(33)$ & $-39.548978$
 & $-13.6075270(18)$ & $  -10.6311(33)$ & $  -14.5997(11)$ \\
$ 0.35$ & $   -9.7248(11)$ & $   10.8512(33)$ & $-39.442149$
 & $-13.554376(17)$ & $  -10.8511(34)$ & $  -14.4555(11)$ \\
\end{tabular}
\end{center}
\end{table}
\begin{table}
\caption{Results for Symanzik(continued) without mean filed improvement.}
\label{tab:Sy_2}
\begin{center}
\begin{tabular}{|l|llll|l|}
 $M$ & $z_{+}$(DRED) & $z_{-}$(DRED) & $z_{1}$(DRED) & $z_{2}$(DRED) &
 $z_{PD}$(DRED)\cr 
\hline
$ 0.05$ & $  -40.4456(22)$ & $  -30.7453(44)$ & $  -38.8289(11)$
 & $  -24.2784(89)$ & $  -24.6757(22)$ \\
$ 0.10$ & $  -39.8843(22)$ & $  -31.1922(44)$ & $  -38.4356(11)$
 & $  -25.3974(88)$ & $  -24.8428(22)$ \\
$ 0.15$ & $  -39.4127(22)$ & $  -31.5351(43)$ & $  -38.0998(11)$
 & $  -26.2834(87)$ & $  -24.9643(22)$ \\
$ 0.20$ & $  -38.9962(22)$ & $  -31.8245(45)$ & $  -37.8009(11)$
 & $  -27.0433(90)$ & $  -25.0636(23)$ \\
$ 0.25$ & $  -38.6187(22)$ & $  -32.0816(45)$ & $  -37.5292(11)$
 & $  -27.7236(89)$ & $  -25.1507(22)$ \\
$ 0.30$ & $  -38.2710(22)$ & $  -32.3182(44)$ & $  -37.2789(11)$
 & $  -28.3497(88)$ & $  -25.2308(22)$ \\
$ 0.35$ & $  -37.9472(22)$ & $  -32.5407(45)$ & $  -37.0461(11)$
 & $  -28.9364(90)$ & $  -25.3066(23)$ \\
\end{tabular}
\end{center}
\end{table}

\begin{table}
\caption{Results for Iwasaki' ($c_1=-0.27$, $c_2+c_3=-0.04$) without
 mean filed improvement.} 
\label{tab:RG2}
\begin{center}
\begin{tabular}{|l|ll|l|lll|}
$M$ & $z_2$(DRED) & $z_m$(DRED) & $\Sigma_w$ & 
$z_{V/A}$(DRED) & $z_{S/P}$(DRED) & $z_{T}$(DRED) \\
\hline
$ 0.05$ & $   -5.6134(11)$ & $    3.9887(33)$ & $-27.030057$
 & $-9.4255052(54)$ & $   -3.9887(33)$ & $  -11.2378(11)$ \\
$ 0.10$ & $   -5.5333(11)$ & $    4.3931(33)$ & $-26.877090$
 & $-9.34607970(10)$ & $   -4.3932(33)$ & $  -10.9970(11)$ \\
$ 0.15$ & $   -5.4626(11)$ & $    4.7098(33)$ & $-26.742564$
 & $-9.2760051(89)$ & $   -4.7098(33)$ & $  -10.7981(11)$ \\
$ 0.20$ & $   -5.3988(11)$ & $    4.9788(34)$ & $-26.621656$
 & $ -9.212870(42)$ & $   -4.9787(35)$ & $  -10.6242(11)$ \\
$ 0.25$ & $   -5.3406(11)$ & $    5.2173(34)$ & $-26.511642$
 & $ -9.155340(36)$ & $   -5.2173(34)$ & $  -10.4680(11)$ \\
$ 0.30$ & $   -5.2871(11)$ & $    5.4350(33)$ & $-26.410767$
 & $-9.1025605(18)$ & $   -5.4350(33)$ & $  -10.3251(11)$ \\
$ 0.35$ & $   -5.2376(11)$ & $    5.6375(33)$ & $-26.317811$
 & $ -9.053846(17)$ & $   -5.6375(34)$ & $  -10.1926(11)$ \\
\end{tabular}
\end{center}
\end{table}
\begin{table}
\caption{Results for Iwasaki'(continued) without mean filed improvement.}
\label{tab:RG2_2}
\begin{center}
\begin{tabular}{|l|llll|l|}
 $M$ & $z_{+}$(DRED) & $z_{-}$(DRED) & $z_{1}$(DRED) & $z_{2}$(DRED) &
 $z_{PD}$(DRED)\cr 
\hline
$ 0.05$ & $  -28.7592(22)$ & $  -17.8856(44)$ & $  -26.9470(11)$
 & $  -10.6365(88)$ & $  -15.2265(22)$ \\
$ 0.10$ & $  -28.2248(22)$ & $  -18.3190(44)$ & $  -26.5738(11)$
 & $  -11.7151(88)$ & $  -15.3902(22)$ \\
$ 0.15$ & $  -27.7801(22)$ & $  -18.6478(43)$ & $  -26.2581(11)$
 & $  -12.5595(87)$ & $  -15.5079(22)$ \\
$ 0.20$ & $  -27.3904(22)$ & $  -18.9221(47)$ & $  -25.9790(10)$
 & $  -13.2766(92)$ & $  -15.6030(24)$ \\
$ 0.25$ & $  -27.0396(22)$ & $  -19.1635(46)$ & $  -25.7269(10)$
 & $  -13.9128(92)$ & $  -15.6853(23)$ \\
$ 0.30$ & $  -26.7185(22)$ & $  -19.3835(44)$ & $  -25.4960(11)$
 & $  -14.4934(88)$ & $  -15.7601(22)$ \\
$ 0.35$ & $  -26.4212(22)$ & $  -19.5884(45)$ & $  -25.2824(11)$
 & $  -15.0332(90)$ & $  -15.8301(23)$ \\
\end{tabular}
\end{center}
\end{table}

\begin{table}
\caption{Results for Wilson ($c_1=-0.252$, $c_2+c_3=-0.17$) without mean
 filed improvement.} 
\label{tab:Wil}
\begin{center}
\begin{tabular}{|l|ll|l|lll|}
$M$ & $z_2$(DRED) & $z_m$(DRED) & $\Sigma_w$ & 
$z_{V/A}$(DRED) & $z_{S/P}$(DRED) & $z_{T}$(DRED) \\
\hline
$ 0.05$ & $   -4.6999(11)$ & $    2.9518(33)$ & $-24.507204$
 & $-8.5112878(54)$ & $   -2.9518(33)$ & $  -10.3644(11)$ \\
$ 0.10$ & $   -4.6207(11)$ & $    3.3530(33)$ & $-24.357346$
 & $-8.432682000(89)$ & $   -3.3530(33)$ & $  -10.1259(11)$ \\
$ 0.15$ & $   -4.5508(11)$ & $    3.6663(33)$ & $-24.225850$
 & $-8.3634038(88)$ & $   -3.6663(33)$ & $   -9.9291(11)$ \\
$ 0.20$ & $   -4.4878(11)$ & $    3.9318(34)$ & $-24.107889$
 & $ -8.301041(42)$ & $   -3.9318(35)$ & $   -9.7575(11)$ \\
$ 0.25$ & $   -4.4303(11)$ & $    4.1669(34)$ & $-24.000738$
 & $ -8.244258(36)$ & $   -4.1668(34)$ & $   -9.6034(11)$ \\
$ 0.30$ & $   -4.3775(11)$ & $    4.3809(33)$ & $-23.902639$
 & $-8.1922005(19)$ & $   -4.3810(33)$ & $   -9.4626(11)$ \\
$ 0.35$ & $   -4.3288(11)$ & $    4.5797(33)$ & $-23.812370$
 & $ -8.144181(17)$ & $   -4.5797(34)$ & $   -9.3323(11)$ \\
\end{tabular}
\end{center}
\end{table}
\begin{table}
\caption{Results for Wilson(continued) without mean filed improvement.}
\label{tab:Wil_2}
\begin{center}
\begin{tabular}{|l|llll|l|}
 $M$ & $z_{+}$(DRED) & $z_{-}$(DRED) & $z_{1}$(DRED) & $z_{2}$(DRED) &
 $z_{PD}$(DRED)\cr 
\hline
$ 0.05$ & $  -26.4031(22)$ & $  -15.2842(44)$ & $  -24.5499(11)$
 & $   -7.8715(88)$ & $  -13.3163(22)$ \\
$ 0.10$ & $  -25.8736(22)$ & $  -15.7143(44)$ & $  -24.1804(11)$
 & $   -8.9414(88)$ & $  -13.4789(22)$ \\
$ 0.15$ & $  -25.4338(22)$ & $  -16.0396(43)$ & $  -23.8681(11)$
 & $   -9.7768(87)$ & $  -13.5954(22)$ \\
$ 0.20$ & $  -25.0489(22)$ & $  -16.3104(47)$ & $  -23.5925(10)$
 & $  -10.4847(92)$ & $  -13.6892(24)$ \\
$ 0.25$ & $  -24.7030(22)$ & $  -16.5481(46)$ & $  -23.3438(10)$
 & $  -11.1116(92)$ & $  -13.7702(23)$ \\
$ 0.30$ & $  -24.3867(22)$ & $  -16.7642(44)$ & $  -23.1163(11)$
 & $  -11.6826(88)$ & $  -13.8436(22)$ \\
$ 0.35$ & $  -24.0941(22)$ & $  -16.9652(45)$ & $  -22.9060(11)$
 & $  -12.2125(90)$ & $  -13.9120(23)$ \\
\end{tabular}
\end{center}
\end{table}

\begin{table}
\caption{Values of $z_{B_K}$ and $z_{B_P}$ for
Plaquette, Iwasaki and DBW2.}
\label{tab:BK1}
\begin{center}
\begin{tabular}{|l|ll|ll|ll|}
& \multicolumn{2}{c|}{Plaquette} & \multicolumn{2}{c|}{Iwasaki} &
 \multicolumn{2}{c|}{DBW2} \\ 
$M$ & $z_{B_K}$(DRED)&$z_{B_P}$(DRED)&$z_{B_K}$(DRED)&$z_{B_P}$(DRED)
&$z_{B_K}$(DRED)&$z_{B_P}$(DRED)\\
 \hline
$ 0.05$ & $   -3.0209(23)$ & $   -15.105(12)$ & $   -3.6821(24)$
 & $   -18.410(12)$ & $   -4.6894(23)$ & $   -23.447(12)$ \\
$ 0.10$ & $   -2.6775(21)$ & $   -13.388(11)$ & $   -3.3614(22)$
 & $   -16.807(11)$ & $   -4.4049(22)$ & $   -22.024(11)$ \\
$ 0.15$ & $   -2.3987(22)$ & $   -11.994(11)$ & $   -3.1057(22)$
 & $   -15.528(11)$ & $   -4.1851(22)$ & $   -20.926(11)$ \\
$ 0.20$ & $   -2.1559(22)$ & $   -10.780(11)$ & $   -2.8864(23)$
 & $   -14.432(11)$ & $   -4.0017(22)$ & $   -20.009(11)$ \\
$ 0.25$ & $   -1.9367(23)$ & $    -9.684(11)$ & $   -2.6911(22)$
 & $   -13.456(11)$ & $   -3.8422(22)$ & $   -19.211(11)$ \\
$ 0.30$ & $   -1.7341(22)$ & $    -8.670(11)$ & $   -2.5129(22)$
 & $   -12.565(11)$ & $   -3.6998(22)$ & $   -18.499(11)$ \\
\end{tabular}
\end{center}
\end{table}
\begin{table}
\caption{Values of $z_{B_K}$ and $z_{B_P}$ for
Symanzik, Iwasaki' and Wilson.}
\label{tab:BK2}
\begin{center}
\begin{tabular}{|l|ll|ll|ll|}
& \multicolumn{2}{c|}{Symanzik} & \multicolumn{2}{c|}{Iwasaki'} &
 \multicolumn{2}{c|}{Wilson} \\ 
$M$ & $z_{B_K}$(DRED)&$z_{B_P}$(DRED)&$z_{B_K}$(DRED)&$z_{B_P}$(DRED)
&$z_{B_K}$(DRED)&$z_{B_P}$(DRED)\\
 \hline
$ 0.05$ & $   -3.2334(22)$ & $   -16.167(11)$ & $   -3.6245(22)$
 & $   -18.123(11)$ & $   -3.7063(22)$ & $   -18.532(11)$ \\
$ 0.10$ & $   -2.8974(22)$ & $   -14.487(11)$ & $   -3.3019(22)$
 & $   -16.510(11)$ & $   -3.3864(21)$ & $   -16.932(11)$ \\
$ 0.15$ & $   -2.6259(22)$ & $   -13.129(11)$ & $   -3.0441(22)$
 & $   -15.221(11)$ & $   -3.1314(22)$ & $   -15.657(11)$ \\
$ 0.20$ & $   -2.3906(23)$ & $   -11.953(11)$ & $   -2.8228(23)$
 & $   -14.114(11)$ & $   -2.9128(22)$ & $   -14.564(11)$ \\
$ 0.25$ & $   -2.1790(22)$ & $   -10.895(11)$ & $   -2.6254(23)$
 & $   -13.127(11)$ & $   -2.7183(23)$ & $   -13.591(11)$ \\
$ 0.30$ & $   -1.9843(22)$ & $    -9.921(11)$ & $   -2.4450(22)$
 & $   -12.225(11)$ & $   -2.5408(22)$ & $   -12.704(11)$ \\
\end{tabular}
\end{center}
\end{table}


\begin{table}
\caption{Fit parameters $a_i$ and $b_1$ of $P$(plaquette) as a function of 
$g^2$ for the plaquette($c_1=0$) and RG improved($c_1=-0.331$) gauge action.
Fit parameters of $R$(rectangular) for the latter are also included.
}
\label{tab:PMF}
\begin{center}
\begin{tabular}{|lll|llll|l|}
action &$g^2_{\rm max}$&  &$a_1$ & $a_2$ & $a_3$ & $a_4$ & $b_1$ \cr
\hline
Plaq. & 1.07143& $P$ & -1.22304(5) & 0.26789(2) & -0.00252(5) & 0.02313(2) 
& -0.889706 \cr
\hline
Iwasaki & 2.74348 &$P$&  -0.446426(3) & 0.0448185(3) & -0.0014248(2) 
& 0.00020329(5) &-0.3062260 \cr
   & & $R$ & -0.571259(2) & 0.1004102(6) & -0.0074407(3) & 0.00041907(7) &
-0.3023588 \cr
\end{tabular}
\end{center}
\end{table}

\begin{table}
\caption{Values of the perturbative quantities $c_P$, $d_g$,
$c_{Ri}$, $T$, $T-T_{\rm MF}$ and $T_{K_c} - T_{\rm MF}$ 
for the various gauge actions.}
\label{tab:pert}
\begin{center}
\begin{tabular}{|l|ll|ll|lll|ll|l|}
action & $c_1$ & $c_3$ & $c_p$ & $d_g$ & $c_{R1}$& $c_{R2}$ & $c_{R3}$
& $T$ & $T-T_{\rm MF}$ & $T_{K_c}-T_{\rm MF}$ \cr
\hline
Plaquette& 0       & 0    & 1/3    & -0.4682& 0.5747 & 0.5228 & 0.5687 &0.1549334 & 0.0299334 & 0.037858 \cr
Iwasaki  & -0.331  & 0    & 0.1401 & 0.1053 & 0.2689 & 0.2223 & 0.2465 &0.0947597 & 0.0421939 &  0.029989\cr
DBW2     & -1.40686& 0    & 0.0511 & 0.5317 & 0.1040 & 0.0818 & 0.0921 &0.0624262 & 0.0432682 & -0.055640\cr
Symanzik & -1/12   & 0    & 0.2440 & -0.2361& 0.4417 & 0.3846 & 0.4215 &0.1282908 & 0.0367251 &  0.036492\cr
Iwasaki' & -0.27   & -0.04& 0.1471 &        & 0.2797 & 0.2342 & 0.2611 &0.0973746 & 0.042173 & 0.030965\cr
Wilson   & -0.252  & -0.17& 0.1286 & 0.1196 & 0.2439 & 0.2070 & 0.2352 &0.0916234 & 0.0433809 & 0.029927 \cr
\end{tabular}
\end{center}
\end{table}

\begin{table}
\caption{Values of parameters $a_i$ in the interpolation of $\Sigma_w$
as a function of $w_0=1-\tilde M$ for various gauge actions.}
\label{tab:inter_zw}
\begin{center}
\begin{tabular}{|l|lll|llll|}
action & $32\pi^2 T_{\rm MF}$ & $a_0$&$32\pi^2T_{\rm MF}+a_0$ 
& $a_1$ & $a_2$ &$a_3$ & $a_4$ \\
\hline
Plaquette& 39.4784  & -49.4658 & -9.9874 & -0.6944 & -1.1452 & 0.6262 & -0.7525 \\
Iwasaki  &  16.6019 & -24.5135 & -7.9116 & -0.4815 & -0.8488 & 0.4240 & -0.6429 \\
DBW2     &   6.0506 & -10.4557 & -4.4051 & -0.2676 & -0.5018 & 0.2253 & -0.5031 \\
Symanzik &  28.9190 & -38.6134 & -9.6945 & -0.6187 & -1.0446 & 0.5540 & -0.7151 \\
Iwasaki' &  17.4342 & -25.6212 & -8.1870 & -0.4978 & -0.8724 & 0.4394 & -0.6522 \\
Wilson   &  15.2363 & -23.1406 & -7.9043 & -0.4766 & -0.8385 & 0.4192 & -0.6411 \\
\end{tabular}
\end{center}
\end{table}

\begin{table}
\caption{Values of parameters $a_i$ in the interpolation of $z_2^{\rm MF}$
as a function of $w_0=1-\tilde M$ for various gauge actions.}
\label{tab:inter_z2}
\begin{center}
\begin{tabular}{|l|ll|llll|ll|}
action &\multicolumn{2}{c|}{$8\pi^2 T_{\rm MF}+a_0$} & $a_1$ & $a_2$ &$a_3$ 
& $a_4$ & $a_5$ & $a_6$ \\
\hline
       & DRED & NDR & & & & & & \\
\hline
Plaquette& -3.0098 & -2.0098 & -0.3399 & -0.5657 & 0.3061 & -0.3727 & & \\
Iwasaki  & -0.3442 &  0.6558 & -0.2618 & -0.4624 & 0.2316 & -0.3340 & & \\
DBW2     & 2.0037 & 3.0037 & -0.1402 & -0.3846 & 0.0291 & -0.0280 & 0.1037 & -0.1933 \\
Symanzik & -2.0735 & -1.0735 & -0.3142 & -0.5337 & 0.2815 & -0.3601 & & \\
Iwasaki' & -0.5033 &  0.4968 & -0.2700 & -0.4733 & 0.2393 & -0.3381 & & \\
Wilson   & -0.1504 &  0.8497 & -0.2652 & -0.4638 & 0.2348 & -0.3359 & & \\
\end{tabular}
\end{center}
\end{table}

\begin{table}
\caption{Values of parameters $a_i$ in the interpolation of $z_m^{\rm MF}$
as a function of $w_0=1-\tilde M$ for various gauge actions.}
\label{tab:inter_zm}
\begin{center}
\begin{tabular}{|l|ll|llll|ll|ll|}
action &\multicolumn{2}{c|}{$-8\pi^2 T_{\rm MF}+a_0$} & $a_1$ & $a_2$ &$a_3$ 
& $a_4$ & $a_5$ & $a_6$ & $a_7$ & $a_8$ \\
\hline
       & DRED & NDR  & & & & & & & & \\
\hline
Plaquette& 7.7341  & 8.7341  & -4.2837 & 1.2934 & -2.8593 & 1.1943 & & &&\\
Iwasaki & 3.2278 & 4.2278 & -3.7023 & 1.1366 & -0.4148 & 0.0823 & -2.0294 & 0.7181 &&\\
DBW2    & -1.7148& -0.7148& -2.3699& 0.5504& -1.4802& 0.5480& 1.2331& -0.4451& -2.0512& 0.6116\\
Symanzik & 6.1997  & 7.1997  & -3.9326 & 1.1268 & -2.7623 & 1.1360 & & &&\\
Iwasaki' & 3.5447 & 4.5447 & -3.7840 & 1.1757 & -0.4348 & 0.0914 & -2.0383 & 0.7237 &&\\
Wilson   & 2.9771 & 3.9771 & -3.6712 & 1.1320 & -0.4158 & 0.0839 & -2.0303 & 0.7184 &&\\
\end{tabular}
\end{center}
\end{table}

\begin{table}
\caption{Values of parameters $a_i$ in the interpolation of $z_\Gamma^{\rm MF}$
as a function of $w_0=1-\tilde M$ for various gauge actions.}
\label{tab:inter_zbi}
\begin{center}
\begin{tabular}{|ll|ll|llll|ll|ll|}
action&$\Gamma$ &\multicolumn{2}{c|}{$8\pi^2 T_{\rm MF}+a_0$} 
& $a_1$ & $a_2$ &$a_3$ & $a_4$ & $a_5$ & $a_6$ & $a_7$ & $a_8$ \\
\hline
      &  & DRED & NDR & &&&&&&&\\
\hline
Plaquette& VA & -6.8787 & -7.3787 & -0.2734 & -0.6126 &  0.3087 & -0.3509 &&&&\\
         & SP & -7.7341 &-8.7341 &4.2837 &-1.2934 &2.8593 &-1.1943 & & & &\\
         & T  & -6.5954 & -5.5954 & -1.8023 & -0.3587 & -0.5198 & -0.1198 &&&&\\
Iwasaki  & VA & -4.1714 & -4.6714 & -0.2264 & -0.4920 &  0.2233 & -0.3074 &&&&\\
         & SP & -3.2278 &-4.2278 &3.7023 &-1.1366 &0.4148 &-0.0823 &2.0294  &-0.7181 &&\\
         & T  & -4.4831 & -3.4831 & -1.4031 & -0.3538 & -0.5281 & -0.1163 &&&&\\
DBW2     & VA & -1.8047 & -2.3047 & -0.1441 & -0.3251 &  0.1256 & -0.2512 &&&&\\
         & SP & 1.7148 &0.7148 &2.3699 &-0.5504 &1.4802 &-0.5480 &-1.2331 &0.4451 &2.0512 &-0.6116 \\
         & T  & -2.9755 & -1.9755 & -0.9049 & -0.2829 & -0.5213 & -0.1143 &&&&\\
Symanzik & VA & -5.9247 & -6.4247 & -0.2582 & -0.5351 &  0.2744 & -0.3822 &&&&\\
         & SP & -6.1997 &-7.1997 &3.9326 &-1.1268 &2.7623 &-1.1360 & &  && \\
         & T  & -5.8342 & -4.8342 & -1.6704 & -0.3620 & -0.5230 & -0.1181 &&&&\\
Iwasaki' & VA & -4.3349 & -4.8349 & -0.2408 & -0.4888 &  0.2500 & -0.3438 &&&&\\
         & SP & -3.5447 &-4.5447 &3.7840 &-1.1757 &0.4348 &-0.0914 &2.0383 &-0.7237 &&\\
         & T  & -4.5939 & -3.5939 & -1.4375 & -0.3578 & -0.5263 & -0.1169 &&&&\\
Wilson   & VA & -3.9808 & -4.4808 & -0.2365 & -0.4791 &  0.2454 & -0.3416 &&&&\\
         & SP & -2.9771 &-3.9771 &3.6712 &-1.1320 &0.4158 &-0.0839 &2.0303 &-0.7184 &&\\
         & T  & -4.3105 & -3.3105 & -1.3949 & -0.3590 & -0.5231 & -0.1186 &&&&\\
\end{tabular}
\end{center}
\end{table}

\begin{table}
\caption{Values of parameters $a_i$ in the interpolation of $z_I^{\rm MF}$
as a function of $w_0=1-\tilde M$ for various gauge actions.}
\label{tab:inter_z4q}
\begin{center}
\begin{tabular}{|ll|ll|llll|llll|}
action&$z_I$ &\multicolumn{2}{c|}{$16\pi^2 C_F T_{\rm MF}+a_0$} 
& $a_1$ & $a_2$ &$a_3$ & $a_4$ & $a_5$ & $a_6$ & $a_7$ & $a_8$ \\
\hline
       & & DRED & NDR &&&&&&&&\\
\hline
Plaquette&$z_+$ & -17.7755 & -20.7755 & -3.7921 & -1.1211 & -0.8227 &  0.4906 & & & & \\
         &$z_-$ & -19.5095 & -17.5095 &  5.3566 & -2.4018 &  4.2146 & -2.2148 & & & & \\
         &$z_1$ & -18.0616 & -17.2283 & -2.2711 & -1.3464 &  0.0227 & -0.7684 & & & & \\
         &$z_2$ & -20.6543 & -23.3210 & 11.4490 & -3.3084 &  7.5830 & -3.3158 & & & & \\
Iwasaki  &$z_+$ & -11.7462 & -14.7462 & -2.9621 & -1.0310 & -0.8968 & -0.4536 & & & & \\
         &$z_-$ &  -9.9090 &  -7.9090 &  4.0754 & -1.6247 &  3.6731 & -1.9229 & & & & \\
\end{tabular}
\end{center}
\end{table}

\begin{table}
\caption{Values of parameters $a_i$ in the interpolation of $z_{PD}^{\rm MF}$
as a function of $w_0=1-\tilde M$ for various gauge actions.}
\label{tab:inter_z3q}
\begin{center}
\begin{tabular}{|l|ll|llll|ll|}
action &\multicolumn{2}{c|}{$12\pi^2 C_F T_{\rm MF}+a_0$} 
& $a_1$ & $a_2$ &$a_3$ & $a_4$ & $a_5$ & $a_6$ \\
\hline
       & DRED & NDR & & & & & & \\
\hline
Plaquette& -14.3397 & -13.6730 & 2.4915 & -1.6054 & 2.3230 & -1.3567  & & \\
Iwasaki  &  -7.7345 &  -7.0678 & 1.8824 & -1.1368 & 1.9949 & -1.1805  & & \\
DBW2     &  -1.2816 &  -0.6149 & 1.2163 & -0.6239 & 1.5858 & -0.9385  & & \\
Symanzik & -12.0114 &   -11.3447  & 2.5984 & -1.8575 & 0.5655 & -0.1570 & 1.5948 & -0.8303 \\
Iwasaki' &  -8.1512 &  -7.4845 & 1.9133 & -1.2459 & 2.0505 & -1.1324  & & \\
Wilson   &  -7.3006 &  -6.6339 & 1.8448 & -1.2045 & 2.0260 & -1.1205  & & \\
\end{tabular}
\end{center}
\end{table}

\begin{table}
\caption{Values of parameters $a_i$ in the interpolation of
 $z_{BK,P}^{\rm MF}$(DRED) 
as a function of $w_0=1-\tilde M$ for various gauge actions.}
\label{tab:inter_zbk}
\begin{center}
\begin{tabular}{|ll|llllllllll|}
action & & $a_0$ & $a_1$ & $a_2$ &$a_3$ & $a_4$ & $a_5$ & $a_6$ & $a_7$ & $a_8$ & $a_9$ \\
\hline
Plaquette&$B_K$&  0.5620& -3.2600& 0.5721& -0.6662& 0.3607& -1.4980& -0.1876& 2.1054& 0.3137& -1.7541\\
         &$B_P$&    2.8090& -16.2960& 2.8702& -3.3753& 1.7750& -7.3424& -0.9065& 10.3304& 1.5573& -8.6802\\

Iwasaki  &$B_K$& -0.6274& -2.5482& 0.3314& -0.5474& 0.3056& -1.4760& -0.2045& 2.1211& 0.3091& -1.7546\\
          &$B_P$&   -3.1382& -12.7364& 1.6692& -2.7886& 1.4842& -7.2027& -0.9620& 10.3672& 1.5174& -8.6625\\
 DBW2     &$B_K$& -2.3449& -1.7033& 0.1284& -0.3989& 0.2266& -1.4279& -0.2110& 2.1224& 0.2867& -1.7398\\
          &$B_P$& -11.7253& -8.5119& 0.6501& -2.0474& 1.1082& -6.9618& -1.0262& 10.3788& 1.4224& -8.5939\\
 Symanzik &$B_K$&   0.1755& -3.0186& 0.4835& -0.6220& 0.3433& -1.5024& -0.1967& 2.1251& 0.3153& -1.7602\\
          &$B_P$&  0.8768& -15.0883& 2.4285& -3.1641& 1.6727& -7.3202& -0.9141& 10.3648& 1.5397& -8.6818\\
 Iwasaki' &$B_K$&  -0.5250& -2.6062& 0.3490& -0.5562& 0.3107& -1.4839& -0.2043& 2.1274& 0.3111& -1.7577\\
          &$B_P$&  -2.6257& -13.0268& 1.7564& -2.8306& 1.5053& -7.2390& -0.9447& 10.3860& 1.5150& -8.6712\\
 Wilson   &$B_K$&  -0.6673& -2.5283& 0.3268& -0.5469& 0.3049& -1.4727& -0.2047& 2.1159& 0.3097& -1.7515\\
          &$B_P$&  -3.3370& -12.6378& 1.6453& -2.7764& 1.4763& -7.2135& -0.9484& 10.3706& 1.5099& -8.6595\\
\end{tabular}
\end{center}
\end{table}

\begin{table}
\caption{Top: Values of $z_{i}^{\rm pen}$(DRED) as a function of $M$,
together with interpolated values and relative errors.
Bottom: Values of parameters $a_i$ and $b_i$ in the interpolation of
$z_{i}^{\rm pen}$(DRED) as a function of $w_0 = 1-\tilde M$.}
\label{tab:pen}
\begin{center}
\begin{tabular}{|l|c|c|c||l|c|c|c|}
$M$ & $z_{i}^{\rm pen}$(DRED) & interpolation & relative error &
$M$ & $z_{i}^{\rm pen}$(DRED) & interpolation & relative error \\
\hline
0.05 &    5.729(1) &  5.734 &  0.001& 1.05 &    0.980(1) &  0.994 &  0.014  \\
0.10 &    1.544(1) &  1.514 &  0.019& 1.10 &    1.178(2) &  1.191 &  0.011  \\
0.15 &    0.355(1) &  0.375 &  0.055& 1.15 &    1.400(1) &  1.410 &  0.007  \\
0.20 &   -0.131(1) & -0.105 & -0.195& 1.20 &    1.648(1) &  1.654 &  0.004  \\
0.25 &   -0.352(1) & -0.335 & -0.048& 1.25 &    1.928(1) &  1.929 &  0.001  \\
0.30 &   -0.448(1) & -0.443 & -0.012& 1.30 &    2.246(1) &  2.243 &  0.001  \\
0.35 &   -0.476(1) & -0.481 & -0.010& 1.35 &    2.614(1) &  2.605 &  0.003  \\
0.40 &   -0.464(1) & -0.475 & -0.024& 1.40 &    3.043(1) &  3.030 &  0.004  \\
0.45 &   -0.425(1) & -0.440 & -0.036& 1.45 &    3.551(1) &  3.536 &  0.004  \\
0.50 &   -0.367(1) & -0.384 & -0.045& 1.50 &    4.163(1) &  4.147 &  0.004  \\
0.55 &   -0.296(1) & -0.311 & -0.050& 1.55 &    4.916(1) &  4.902 &  0.003  \\
0.60 &   -0.213(1) & -0.225 & -0.057& 1.60 &    5.865(1) &  5.856 &  0.002  \\
0.65 &   -0.120(1) & -0.128 & -0.070& 1.65 &    7.097(1) &  7.095 &  0.000  \\
0.70 &   -0.019(1) & -0.022 & -0.163& 1.70 &    8.759(1) &  8.768 &  0.001  \\
0.75 &    0.092(1) &  0.093 &  0.013& 1.75 &   11.120(1) & 11.138 &  0.002  \\
0.80 &    0.211(1) &  0.217 &  0.030& 1.80 &   14.721(1) & 14.742 &  0.001  \\
0.85 &    0.341(1) &  0.351 &  0.028& 1.85 &   20.841(1) & 20.844 &  0.000  \\
0.90 &    0.480(1) &  0.493 &  0.028& 1.90 &   33.357(1) & 33.339 &  0.001  \\
0.95 &    0.632(1) &  0.647 &  0.024& 1.95 &  71.853(22) & 72.947 &  0.015  \\
1.00 &    0.797(1) &  0.813 &  0.020&      &             &        &         \\
\end{tabular}
\end{center}
\begin{center}
\begin{tabular}{|l|lllll|llll|}
Action & $a_0$ & $a_1$ & $a_2$ &$a_3$ &$a_4$ & $b_1$ & $b_2$ &$b_3$ &$b_4$ \\
\hline
All& 0.8132 & -3.4681 & 2.3612 & -0.8213 & 2.0186 & -0.0127 & -0.6956 & 0.0054 & -0.3216 \\
\end{tabular}
\end{center}
\end{table}

\begin{table}
\caption{Values of parameters $a_i$ in the interpolation of $z_{ij}^{\rm MF}$(DRED)
as a function of $w_0=1-\tilde M$ for various gauge actions.}
\label{tab:delS}
\begin{center}
\begin{tabular}{|ll|c|llll|ll|ll|}
action&$z_{ii}$ & $16\pi^2 C_F T_{\rm MF}+a_0$
& $a_1$ & $a_2$ &$a_3$ & $a_4$ & $a_5$ & $a_6$ & $a_7$ & $a_8$ \\
\hline
Plaquette& $z_{11}$ &-18.6311 &  0.7719 & -1.8154 &  1.7127 & -1.3019 &&&& \\
         & $z_{55}$ &-18.0616 & -2.2711 & -1.3464 &  0.0227 & -0.7684 &&&&\\
         & $z_{66}$ &-20.6543 & 11.4490 & -3.3084 &  7.5830 & -3.3158 &&&&\\
Iwasaki  & $z_{11}$ &-10.8167 &  0.5467 & -1.3795 &  1.4045 & -1.1395 &&&&\\
         & $z_{55}$ &-11.4375 & -1.7924 & -1.1399 & -0.1300 & -0.6906 &&&&\\
         & $z_{66}$ & -8.6739 &  8.7611 & -2.0690 &  6.7288 & -2.8579 &&&&\\
DBW2     & $z_{11}$  & -3.6286 & 0.6247 & -1.1155 & -0.3311 & -0.1293 & 1.3053 & -0.5945 &&\\
         & $z_{55}$ & -5.9855 & -1.1556 & -0.7930 & -0.2895 & -0.5888 &&&&\\
          &$z_{66}$ &  4.5663 & 6.3348& -1.3762& 3.8436& -1.8212& -3.0864& 1.7158& 5.3507& -1.8910\\
Symanzik & $z_{11}$ &-15.8928 &  0.6820 & -1.6663 &  1.6158 & -1.2490 &&&&\\
         & $z_{55}$ &-15.7093 & -2.1204 & -1.2884 & -0.0254 & -0.7338 &&&&\\
         & $z_{66}$ &-16.5657 & 10.5000 & -2.8547 &  7.3568 & -3.1659 &&&&\\
Iwasaki' & $z_{11}$ &-11.3034 &  0.5562 & -1.4186 &  1.4431 & -1.1581 &&&&\\
         & $z_{55}$ &-11.8187 & -1.8391 & -1.1732 & -0.1092 & -0.6892 &&&&\\
         & $z_{66}$ & -9.4850 &  8.9385 & -2.3597 &  6.8666 & -2.7263 &&&&\\
Wilson   & $z_{11}$ &-10.2878 &  0.5290 & -1.3824 &  1.4235 & -1.1480 &&&&\\
         & $z_{55}$ &-10.9454 & -1.7894 & -1.1583 & -0.1136 & -0.6872 &&&&\\
         & $z_{66}$ & -7.9715 &  8.6425 & -2.2482 &  6.7932 & -2.6882 &&&&\\
\end{tabular}
\end{center}
\begin{center}
\begin{tabular}{|ll|lllll|lllll|}
action&$z_{ij}$ &$a_0$ & $a_1$ & $a_2$ &$a_3$ & $a_4$ &
$a_5$ & $a_6$ &$a_7$ &$a_8$ &$a_9$ \\
\hline
Plaquette& $z_{12}$  & 0.8427& -4.8888& 0.8613& -1.0126& 0.5315& -2.2025& -0.2703& 3.0984& 0.4663& -2.6036\\
Iwasaki   & $z_{12}$ & -0.9415& -3.8209& 0.5009& -0.8372& 0.4442& -2.1582& -0.2866& 3.1060& 0.4541& -2.5966\\
DBW2      & $z_{12}$ & -3.5176& -2.5536& 0.1952& -0.6143& 0.3317& -2.0881& -0.3064& 3.1126& 0.4259& -2.5776\\
Symanzik  & $z_{12}$ &  0.2631& -4.5265& 0.7285& -0.9498& 0.5018& -2.1934& -0.2742& 3.1050& 0.4618& -2.6022\\
Iwasaki'  & $z_{12}$ & -0.7877& -3.9080& 0.5269& -0.8502& 0.4517& -2.1672& -0.2835& 3.1087& 0.4545& -2.5978\\
Wilson    & $z_{12}$ & -1.0011& -3.7913& 0.4935& -0.8332& 0.4434& -2.1629& -0.2853& 3.1093& 0.4533& -2.5969\\
\end{tabular}
\end{center}
\end{table}


\begin{thebibliography}{99}
\bibitem{kaneko}
We would like to thank Dr. T. Kaneko for providing us
the results of these fits.
\end{thebibliography}
\end{document}